\documentclass[useAMS,usenatbib]{mn2e}

\usepackage{epsfig}
\usepackage{longtable}
\usepackage{times} 
\newcommand{\sw}{$Swift$}

\def \igr17544 {\mbox{IGR~J17544$-$2619}}
\def \xte17391 {\mbox{XTE~J1739$-$302}}
\def \src16479 {\mbox{IGR~J16479$-$4514}}

\def \inte {{\em INTEGRAL}}
\def \rxte {{\em RXTE}}
\def \sax {{\em BeppoSAX}}
\def \suzaku {{\em Suzaku}}
\def \sw {{\em Swift}}
\def \xmm {{\em XMM--Newton}}

\def \hcm {\hbox {\ifmmode $ atom cm$^{-2}\else atom cm$^{-2}$\fi}}

\def \sax{{\it BeppoSAX}}
\def \ATel {Astron.\ Tel.}
\def \apj {ApJ}
\def \apjl {ApJL}
\def \aap {A\&A}

\def \mnras {MNRAS}

\title[SFXTs with {\it Swift}: two years of monitoring]{Two years of monitoring Supergiant Fast X--ray Transients with \emph{Swift} }

\author[P.\ Romano et al.]{P.\ Romano$^{1}$, V.\ La Parola$^{1}$, 
S.\ Vercellone$^{1}$, G.\ Cusumano$^{1}$, L.~ Sidoli$^{2}$, 
H.A.\ Krimm$^{3,4}$, 
\newauthor  C.\ Pagani$^{5,6}$, P.\ Esposito$^{7,8}$, E.A.~Hoversten$^{5}$, J.A.~Kennea$^{5}$, K.L.~Page$^6$, 
 D.N.~Burrows$^{5}$, 
\newauthor  N.~Gehrels$^{4}$ \\
$^{1}$INAF, Istituto di Astrofisica Spaziale e Fisica Cosmica,
        Via U.\ La Malfa 153, I-90146 Palermo, Italy\\
$^{2}$INAF, Istituto di Astrofisica Spaziale e Fisica Cosmica,
	Via E.\ Bassini 15,   I-20133 Milano,  Italy\\
$^{3}$NASA/Goddard Space Flight Center, Greenbelt, MD 20771, USA\\
$^{4}$Universities Space Research Association, Columbia, MD, USA \\
$^{5}$Department of Astronomy and Astrophysics, Pennsylvania State 
             University, University Park, PA 16802, USA\\
$^{6}$Department of Physics \& Astronomy, University of Leicester, LE1 7RH, UK\\
$^{7}$INAF, Osservatorio Astronomico di Cagliari, localit\`a Poggio dei Pini, strada 54, I-09012 Capoterra, Italy\\
$^{8}$Istituto Nazionale di Fisica Nucleare, sezione di Pavia, via A.~Bassi 6, I-27100 Pavia, Italy\\
}

\begin{document}

\date{Accepted 2010 August 18.  Received 2010 August 17; in original form 2010 March 17}

\pagerange{\pageref{firstpage}--\pageref{lastpage}} \pubyear{2010}

\maketitle

\label{firstpage}

\begin{abstract}
We present results based on two years of intense {\em Swift} 
monitoring of three supergiant fast X--ray transients (SFXTs), 
IGR~J16479$-$4514,  XTE~J1739$-$302, and IGR~J17544$-$2619,
which we started in October 2007.  
Our out-of-outburst intensity-based X--ray (0.3--10\,keV) spectroscopy yields  
absorbed power laws characterized by hard photon indices ($\Gamma$$\sim$1--$2$). 
The broad-band (0.3--150\,keV) spectra of these sources, obtained while they were 
undergoing new outbursts observed during the second year of monitoring, 
can be fit well with models typically used to describe 
the X--ray emission from accreting neutron stars in high-mass X--ray binaries. 
We obtain an assessment of how long each source 
spends in each state using a systematic monitoring with a sensitive instrument. 
By considering our monitoring as a casual sampling of the X--ray light curves,
we can infer that the time these sources spend in bright outbursts  
is between 3 and 5\,\% of the total. 
The most probable X-ray flux for these sources is 
$\sim 1$--$2\times10^{-11}$ erg cm$^{-2}$ s$^{-1}$ 
(2--10\,keV, unabsorbed), corresponding to luminosities in the order of a few $10^{33}$ 
to a few $10^{34}$ erg s$^{-1}$ (two orders of magnitude lower than the bright outbursts). 
In particular, the duty-cycle of {\it inactivity} is $\sim 19, 39, 55$\,\%  
($\sim 5$\,\% uncertainty), for IGR~J16479$-$4514, XTE~J1739$-$302, and IGR~J17544$-$2619,  
respectively. We present a complete list of BAT on-board detections, which 
further confirms the continued activity of these sources. 
This demonstrates that true quiescence is a rare state, 
and that these transients accrete matter throughout their life at 
different rates. 
Variability in the X--ray flux is observed at all timescales and 
intensity ranges we can probe. Superimposed on the day-to-day variability 
is intra-day flaring which involves flux variations up to one order of 
magnitude that can occur down to timescales as short as $\sim 1$\,ks, and which 
can be naturally explained by the accretion of single clumps 
composing the donor wind with masses $M_{\rm cl} \sim 0.3$--$2\times10^{19}$ g.
Thanks to the \sw\ observations, the general picture we obtain is that, 
despite individual differences, common X--ray characteristics of this class 
are now well defined, such as outburst lengths well in excess of hours, with a
multiple peaked structure, and a high dynamic range (including bright outbursts),  
up to $\sim4$ orders of magnitude. 
\end{abstract}

\begin{keywords}
X-rays: binaries -- X-rays: individual: IGR~J16479$-$4514, XTE~J1739$-$302, IGR~J17544$-$2619. 

\noindent
Facility: {\it Swift}

\end{keywords}


\setcounter{table}{3} 
 \begin{table*}
 \begin{center}
 \caption{Summary of the {\it Swift}/XRT monitoring campaign.\label{sfxt6:tab:campaign} }
 \begin{tabular}{lrrrrlll}
 \hline
 \noalign{\smallskip}
Name &Campaign &Campaign &N$^{\mathrm{a}}$ &Exposure$^{\mathrm{b}}$ &Outburst$^{\mathrm{c}}$ &BAT &References \\
     &       Start         &End             & &       & Dates   & Trigger   & \\
     &       (yyyy-mm-dd)  &(yyyy-mm-dd)    & & (ks)  & (yyyy-mm-dd) &   & \\
  \noalign{\smallskip}
 \hline
 \noalign{\smallskip}
IGR~J16479$-$4514 & {\it 2007-10-26} & {\it 2008-10-25}& {\it 70}&  {\it 75.2} & {\it 2008-03-19}& {\it 306829} &{\it \citet{Romano2008:sfxts_paperII} }\\
                  &                  &                 &         &             & {\it 2008-05-21}& {\it 312068} & \\ 
                  & 2009-01-29       & 2009-10-25      &  74     &  85.7       & 2009-01-29      & 341452       & \citet{Romano2009:atel1920,LaParola2009:atel1929}  \\ 
  \noalign{\smallskip}
XTE~J1739$-$302   & {\it 2007-10-27} & {\it 2008-10-31}& {\it 95}&  {\it 116.1}& {\it 2008-04-08}& {\it 308797} & {\it \citet{Romano2008:atel1466,Sidoli2009:sfxts_paperIII} }\\
                  &	             &                 &         &             & {\it 2008-08-13}& {\it 319963} & 
                                                                                                       {\it \citet{Romano2008:atel1659}, \citet{Sidoli2009:sfxts_paperIV} }\\
                  & 2009-02-21       & 2009-11-01      & 89      &  89.6       & 2009-03-10      & 346069       & \citet{Romano2009:atel1961}, this work  \\ 
  \noalign{\smallskip}
IGR~J17544$-$2619 & {\it 2007-10-2}8 & {\it 2008-10-31}& {\it 77}&  {\it 74.8} & {\it 2007-11-08}&  BTM         & {\it \citet{Krimm2007:ATel1265} }\\
  		  &                  &                 &         &             & {\it 2008-03-31}& {\it 308224} & {\it \citet{Sidoli2008:atel1454,Sidoli2009:sfxts_paperIII} }\\
  		  &   	             &                 &         &             & {\it 2008-09-04}&  XM          & {\it \citet{Romano2008:atel1697,Sidoli2009:sfxts_paperIV} }  \\
                  & 2009-02-21       & 2009-11-03      & 65      &  68.0       & 2009-03-15      &  BTM         & \citet{Krimm2009:atel1971} \\ 
                  &   	             &                 &         &             & 2009-06-06      & 354221       & \citet{Romano2009:atel2069} \\
  \noalign{\smallskip}
 \hline
 \noalign{\smallskip}
First year        &                 &                  &{\it 330}& {\it 362.6} &                 &              & \\  
Second year       &                 &                  &228      &  243.3      &                 &              & \\  
Total             &                 &                  &558      &  605.9      &                 &              & \\  
  \noalign{\smallskip}
  \hline
  \end{tabular}
  \end{center}
  \begin{list}{}{}
  \item[$^{\mathrm{a}}$]{Number of observations (individual OBSIDs) obtained during the monitoring campaign.}
  \item[$^{\mathrm{b}}$]{\sw/XRT net exposure.}
  \item[$^{\mathrm{c}}$]{BAT trigger dates. BTM=triggered the BAT Transient Monitor; XM=discovered in XRT monitoring data. 
   } \\ 
  {\it Note.} We report the data from the first year in italics.
  \end{list}
  \end{table*}


	\section{Introduction\label{sfxt6:intro}}

Supergiant fast X--ray transients (SFXTs) constitute a new class of High Mass
X--ray Binaries (HMXBs).  Discovered by \inte{} \citep{Sguera2005},  
they are firmly associated with OB supergiant stars via optical spectroscopy, 
and display sporadic X--ray outbursts significantly shorter 
than those of typical Be/X--ray binaries,  characterized  
(as observed by \inte/IBIS) by bright flares 
(peak luminosities of 10$^{36}$--10$^{37}$~erg~s$^{-1}$)
lasting a few hours  \citep{Sguera2005,Negueruela2006}.
The quiescence, which is characterized by a soft spectrum (likely thermal) 
shows a luminosity of $\sim 10^{32}$~erg~s$^{-1}$
so that SFXTs display a dynamic range of 3--5 orders of magnitude. 
Their hard X--ray spectra during outburst resemble those of HMXBs 
hosting accreting neutron stars, with hard power laws below 10\,keV combined 
with high energy cut-offs at $\sim 15$--30~keV, 
sometimes strongly absorbed at soft energies \citep{Walter2006,SidoliPM2006}. 
Therefore, it is tempting to assume that all SFXTs might host a neutron star, 
even if pulse periods have only been measured for a few SFXTs. 
Consensus has not been reached yet on the actual mechanism producing the outbursts,
but it is probably related to either the properties of 
the wind from the supergiant companion 
\citep{zand2005,Walter2007,Negueruela2008,Sidoli2007} or the 
presence of gated mechanisms \citep[][]{Bozzo2008}.

\sw{} is currently the only observatory which, thanks to its 
unique fast-slewing capability and its broad-band energy coverage, 
can catch outbursts from these fast transients  in their very early stages and study them 
panchromatically as they evolve, 
thus providing invaluable information on  the nature 
of the mechanisms producing them. 

Furthermore, thanks to its flexible observing scheduling, which makes a monitoring 
effort cost-effective, \sw{} has given SFXTs the first non serendipitous 
attention through monitoring campaigns that cover all phases of their lives 
with a high sensitivity in the soft X--ray regime, where most SFXTs had not been observed
before \citep{Romano2009:sfxts_paperV}. 

In our previous papers of this series, we described the 
long-term X--ray emission outside the bright outbursts based on the first 4 months of data 
\citep[][Paper I]{Sidoli2008:sfxts_paperI};  
the outbursts of IGR~J16479$-$4514 (\citealt[][]{Romano2008:sfxts_paperII}, Paper II; 
 \citealt[][]{Romano2009:sfxts_paperV}, Paper V); and the prototypical IGR~J17544$-$2619 
and XTE~J17391$-$302  (\citealt{Sidoli2009:sfxts_paperIII}, Paper~III; 
\citealt{Sidoli2009:sfxts_paperIV}, Paper~IV).

In this paper we continue our characterization of  the long term properties 
of our sample as derived from a two-year-long high-sensitivity X--ray coverage. 
We report on the new X--ray Telescope  \citep[XRT, ][]{Burrows2005:XRTmn} 
and the UV/Optical Telescope \citep[UVOT, ][]{Roming2005:UVOTmn}   
data collected in 2009 between January 29 and November 3 
and we exploit the longer baseline for in-depth 
soft X--ray spectral analysis. Furthermore, we show our  results on the outbursts 
caught by the Burst Alert Telescope   \citep[BAT, ][]{Barthelmy2005:BAT} 
 during the second year of our monitoring campaign, as well as the BAT on-board triggers 
registered during 2009.

	\section{Sample and Observations} \label{sfxt6:sample}

During the second year of \sw\ observations, we monitored three targets, 
XTE~J1739--302, IGR~J17544$-$2619, and  IGR~J16479$-$4514. 

\xte17391 {} was discovered in August 1997  by \rxte\ \citep{Smith1998:17391-3021}, 
when it reached a peak flux of 3.6$\times$10$^{-9}$~erg cm$^{-2}$~s$^{-1}$ (2--25 keV)
and has a long history of flaring activity recorded by \inte\ \citep{Sguera2006,Walter2007,Blay2008}  
and by \sw\  \citep{Sidoli2009:sfxts_paperIII,Sidoli2009:sfxts_paperIV}.
Recently, \citet{Drave2010:17391_3021_period} reported the discovery of a $51.47\pm0.02$\,d 
orbital period based on $\sim 12.4$\,Ms of \inte\ data.
The optical counterpart is an O8I star \citep{Negueruela2006}.  

The first recorded flare from \igr17544 {} was observed by \inte\ in 2003 \citep{Sunyaev2003}, 
when the source reached a flux of 160~mCrab (18--25~keV). Several more flares, lasting up to 10 hours, 
were detected by \inte\ in  the following years 
\citep{Grebenev2003:17544-2619,Grebenev2004:17544-2619,Sguera2006,Walter2007,Kuulkers2007} 
with fluxes up to 400~mCrab (20--40 keV) and some were found in archival \sax\ observations 
\citep{zand2004:17544bepposax}. 
Subsequent flares were observed by \sw\  \citep{Krimm2007:ATel1265,Sidoli2009:sfxts_paperIII,Sidoli2009:sfxts_paperIV}, 
and \suzaku\ \citep{Rampy2009:suzaku17544}. 
Recently, \citet{Clark2009:17544-2619period} reported the discovery of a $4.926\pm0.0001$\,d 
orbital period based on the $\sim 4.5$ years of \inte\ data. 
The optical counterpart is an O9Ib star \citep{Pellizza2006}.  

\src16479 {} was discovered by \inte\ in 2003 \citep{Molkov2003:16479-4514}, 
during an outburst that reached the flux level of $\sim 12$\,mCrab (18--25\,keV).
Since then the source has shown frequent flaring activity, recorded by both 
\inte\ \citep[][]{Sguera2005,Sguera2006,Walter2007} and \sw{} 
\citep[][]{Kennea2005:16479-4514,Markwardt2006:16479-4514,Romano2008:sfxts_paperII,
Bozzo2009:16479-4514outburst,Romano2009:sfxts_paperV}, 
which led to its inclusion in the SFXT class. 
The optical counterpart is an O8.5I star \citep{Rahoui2008}. 
Recently, \citet{Bozzo2008:eclipse16479} reported an episode of
sudden obscuration in a long \xmm{} observation obtained after the 2009 March 19 outburst, 
possibly an X--ray eclipse by the supergiant companion. This was later confirmed by 
\citet{Jain2009:16479_period}, who discovered a 3.32\,d orbital period in the first 4 years 
of BAT data and \rxte/ASM data, 
and by \citet{Romano2009:sfxts_paperV} by using XRT observations covering one year. 

We monitored the first two targets as they are generally considered prototypical SFXTs, 
and the latter because of its frequent triggering of the BAT since the beginning of the mission.  
We planned two observations week$^{-1}$ object$^{-1}$ 
(IGR~J16479$-$4514 and IGR~J17544$-$2619) and 3 observations week$^{-1}$ object$^{-1}$ 
(XTE~J1739--302), each 1\,ks long.  
The observation logs are reported in Tables~\ref{sfxt6:tab:alldata16479}, 
\ref{sfxt6:tab:alldata17391}, and \ref{sfxt6:tab:alldata17544}. 
This strategy was chosen to fit within the regular observing schedule
of the main observing targets for \sw,  $\gamma$-ray bursts (GRBs). 
Furthermore, in order to ensure simultaneous narrow field instrument (NFI) 
data, the \sw\ Team enabled 
automatic rapid slews  to these objects following detection of flares 
by the BAT,  in the same fashion as is currently done for GRBs. 
During the campaign we often requested target of opportunity (ToO)
observations whenever one of the sources showed interesting activity,
or following outbursts to better monitor the decay of the XRT light curve, 
thus obtaining a finer sampling of the light curves and allowing us to study all 
phases of the evolution of an outburst. 

During the second year (2009), we collected a total of 228 \sw\ observations as 
part of our program, for a total net XRT exposure of $\sim 243$\,ks accumulated on 
all sources and distributed as shown in Table~\ref{sfxt6:tab:campaign}.

 	 \section{ Data Reduction} \label{sfxt6:dataredu}

The XRT data were uniformly processed with standard procedures 
({\sc xrtpipeline} v0.12.3), filtering and screening criteria by using 
{\sc FTOOLS} in the {\sc Heasoft} package (v.6.7).  
We considered both windowed-timing (WT) and photon-counting (PC) mode data, 
and selected event grades 0--2 and 0--12, respectively 
(\citealt{Burrows2005:XRTmn}).
When appropriate, we corrected for pile-up 
by determining the size of the affected core of the point spread function (PSF)  
by comparing the observed and nominal PSF \citep{vaughan2006:050315mn},
and excluding from the analysis all the events that fell within that
region. 
We used the latest spectral redistribution matrices in CALDB (20091130).  

The BAT data of the outbursts (see Sect.~\ref{sfxt6:outbursts}) 
were analysed using the standard BAT software  
within {\sc FTOOLS}. 
Mask-tagged BAT light curves were created in the standard energy bands,
and rebinned to fulfil at least one of the following conditions, 
achieving a signal-to-noise (S/N) of 5 or bin length of 100\,s. 
Response matrices were generated with {\sc batdrmgen} 
using the latest spectral redistribution matrices. 
For this paper we also considered the BAT Transient Monitor data 
\citep[][]{Krimm2006_atel_BTM,Krimm2008_HEAD_BTM}\footnote{http://swift.gsfc.nasa.gov/docs/swift/results/transients/ }, 
covering the same time interval as the NFI pointed observations.
The data were rebinned to a 4\,d resolution to ensure a larger number of detections
and to closely match the NFI sampling. 

The UVOT observed the 3 targets simultaneously with the XRT.  
The data of \xte17391 {} and \src16479 {} were taken in general with the 
`Filter of the Day' (FoD), i.e.\ the filter chosen for all observations 
to be carried out during a specific day in order to minimize the filter wheel rotation
($u$, $uvw1$, $uvw2$ and $uvm2$), 
while \igr17544 {} was 
only observed with UVOT from 2009-06-06 to 2009-09-30, in the $uvw1$ and $v$ filters,
and in FoD for the remainder of the campaign. 
However, during the outbursts of 2009-03-10 of \xte17391 {} and 
2009-06-06 of \igr17544 {} all filters were used in the typical GRB sequence 
\citep{Roming2005:UVOTmn}. 
The data analysis was performed using the {\sc uvotimsum} and 
{\sc uvotsource} tasks included in the {\sc FTOOLS} software. 
The latter task calculates the magnitude through aperture photometry within
a circular region and applies specific corrections due to the detector
characteristics. The reported magnitudes are on the UVOT photometric 
system described in \citet{Poole2008:UVOTmn}, and 
are not corrected for Galactic extinction. 
At the position of IGR~J16479$-$4514, no detection
was achieved down to a 3$\sigma$ limit of  $u=21.27$ mag.

All quoted uncertainties are given at 90\,\% confidence level for 
one interesting parameter unless otherwise stated. 
The spectral indices are parameterized as  
$F_{\nu} \propto \nu^{-\alpha}$, 
where $F_{\nu}$ (erg cm$^{-2}$ s$^{-1}$ Hz$^{-1}$) is the 
flux density as a function of frequency $\nu$; 
we adopt $\Gamma = \alpha +1$ as the photon index, 
$N(E) \propto E^{-\Gamma}$ (ph cm$^{-2}$ s$^{-1}$ keV$^{-1}$).

\section{Timing}

\begin{figure}
\begin{center}
\vspace{-1.5truecm}
\centerline{\includegraphics[width=9cm,angle=0]{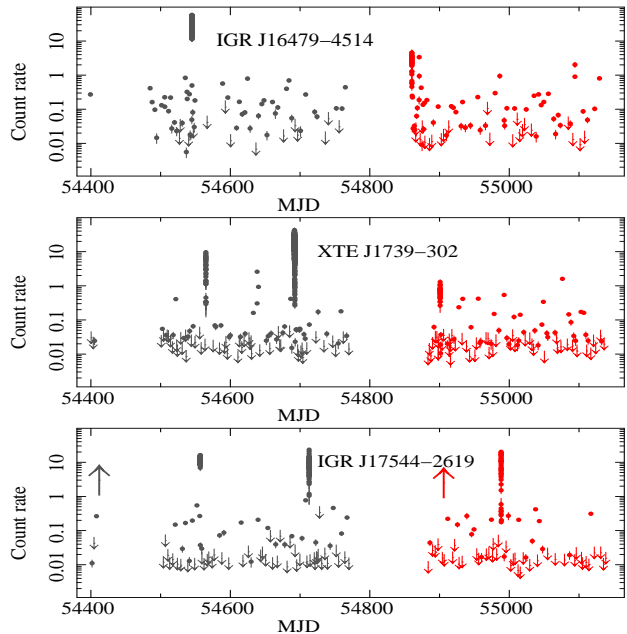}}
\vspace{-2.5truecm}
\caption[XRT light curves]{\sw/XRT (0.2--10\,keV) light curves.  
                The data were collected from 2007 October 26 to 2008 November 15 (first year, grey) 
                and from 2009 January 29 to November 3 (second year, red). 
		The downward-pointing arrows are 3$\sigma$ upper limits. The upward pointing arrows 
                mark flares that triggered the BAT Transient Monitor on MJD 54414 and 54906.  
                Data up to MJD $\sim 54770$ (grey) were published in \citet[][]{Romano2009:sfxts_paperV}. }
		\label{sfxt6:fig:xrtlcvs} 
        \end{center}
        \end{figure}

\begin{figure}
\begin{center}
\vspace{-1.5truecm}
\centerline{\includegraphics[width=9cm,angle=0]{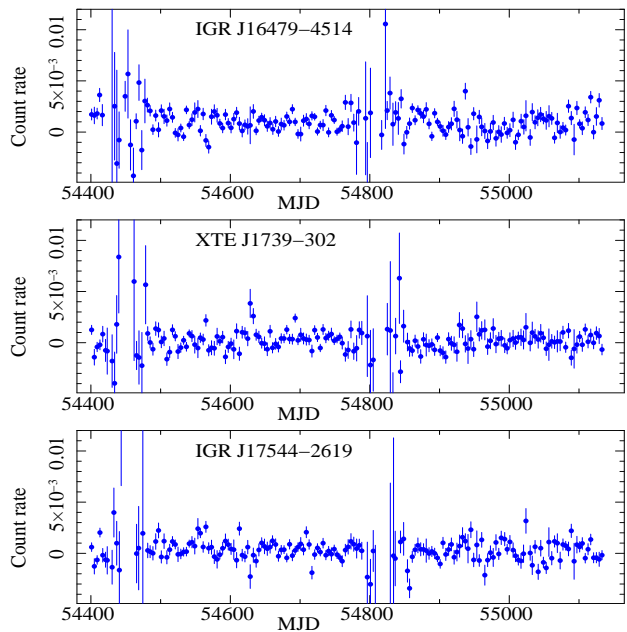}}
\vspace{-2.5truecm}
\caption[XRT light curves]{\sw/BAT Transient Monitor light curves in the 15--50\,keV energy range in units of
counts s$^{-1}$ cm$^{-2}$.  }
		\label{sfxt6:fig:batlcvs} 
        \end{center}
        \end{figure}

        \subsection{XRT light curves and inactivity  duty cycle\label{sfxt6:idc}}

The 0.2--10\,keV XRT light curves collected from 2007 October 26 to  2009 November 3,
are shown in Fig.~\ref{sfxt6:fig:xrtlcvs}.
They are corrected for pile-up, PSF losses, and
vignetting, and background-subtracted. Each point in the
light curves refers to the average flux observed
during each observation performed with XRT; the exceptions are the outbursts 
(listed in Table~\ref{sfxt6:tab:campaign}) where the data were binned to 
include at least 20 source counts per time bin to best represent the 
count rate dynamical range.

 \begin{table}
 \begin{center} 
 \caption{Duty cycle of inactivity of the three SFXTs (2-year campaign).}
 \label{sfxt6:tab:dutycycle}
 \begin{tabular}{lcccc}
 \hline
 \noalign{\smallskip}
Name &$\Delta T_{\Sigma}$  & $P_{\rm short}$ &  IDC  & Rate$_{\Delta T_{\Sigma}}$   \\           
                &(ks) & (\%) &  (\%) &   ($10^{-3}$counts s$^{-1}$)
                   \\  
  \noalign{\smallskip}
 \hline
 \noalign{\smallskip}
IGR~J16479$-$4514   & 29.7 &3  & 19 & $3.1\pm0.5$        \\
XTE~J1739$-$302     & 71.5 &10 & 39 & $4.0\pm0.3$       \\
IGR~J17544$-$2619   & 69.3 &10 & 55 & $2.2\pm0.2$       \\

  \noalign{\smallskip}
  \hline
  \end{tabular}
  \end{center}
  \begin{list}{}{}
\item {Count rates are in units of $10^{-3}$ counts s$^{-1}$ in the 0.2--10\,keV energy band.  
   $\Delta T_{\Sigma}$ is sum of the exposures accumulated in all observations, 
   each in excess of 900\,s, where only a 3$\sigma$ upper limit was achieved;  
   $P_{\rm short}$ is the percentage of time lost to short observations; 
   IDC is the duty cycle of  {\it inactivity}, {\it i.e., }
   the time each source spends  undetected down to a flux limit of 1--3$\times10^{-12}$ erg cm$^{-2}$ s$^{-1}$;
   Rate$_{\Delta T_{\Sigma}}$ is detailed in the text (Sect.~\ref{sfxt6:idc}).
   }
  \end{list}
  \end{table}

One of our goals is to calculate the percentage of time each source spent in each flux state.
We consider our monitoring as a casual sampling of the light curve at a 
resolution of $\sim 3$--4\,d  over a $> 2$ yr baseline. We considered the following three states, 
{\it i)} BAT-detected outburst, 
{\it ii)} intermediate state (all observations yielding a firm detection excluding outburst ones),  
{\it iii)} `non detections' (detections with a significance below 3$\sigma$). 
From the latter state we excluded all observations that had a net exposure below 900\,s
[corresponding to 2--10\,keV flux limits that vary between 1 and 3$\times 10^{-12}$ erg cm$^{-2}$ s$^{-1}$ 
(3$\sigma$), depending on the source, see \citet{Romano2009:sfxts_paperV}]. This was done because 
\sw\ is a GRB-chasing mission and several observations were interrupted by GRB events; therefore 
the consequent non detection may be due to the short exposure, and not exclusively to the source 
being faint. 

The duty cycle of {\it inactivity} is defined  \citep{Romano2009:sfxts_paperV}  as 
the time each source spends {\it undetected} down to a flux limit of 
1--3$\times10^{-12}$ erg cm$^{-2}$ s$^{-1}$,  
\begin{equation}
{\rm IDC}= \Delta T_{\Sigma} / [\Delta T_{\rm tot} \, (1-P_{\rm short}) ] \, ,     
\end{equation}
where  
$\Delta T_{\Sigma}$ is sum of the exposures accumulated in all observations, 
   each in excess of 900\,s, where only a 3$\sigma$ upper limit was achieved 
(Table~\ref{sfxt6:tab:dutycycle}, column 2), 
$\Delta T_{\rm tot}$ is the total exposure accumulated (Table~\ref{sfxt6:tab:campaign}, column 5), and 
$P_{\rm short}$ is the percentage of time lost to short observations  
(exposure $<900$\,s, Table~\ref{sfxt6:tab:dutycycle}, column 2). 
We obtain that ${\rm IDC} = 19, 39, 55$\,\%, 
for IGR~J16479$-$4514, XTE~J1739--302, and IGR~J17544$-$2619, respectively
(Table~\ref{sfxt6:tab:dutycycle}, column 3), with an estimated error 
of $\sim 5\,\%$. 
We note that these values are based on the whole 2-year campaign,
and that the IDC calculated with the observations of 2009 only is $21, 39, 55$\,\%.
Finally, as we can consider our monitoring as a casual sampling of the light curve,
we can also infer that the time these sources spend in bright outbursts  
is between 3 and 5\,\% of the total (estimated error of $\sim 5\,\%$).

       \subsection{BAT transient monitor data and on-board detections}

\begin{table}
 \begin{center}
 \caption{BAT on-board detections in the 15--50\,keV band during 2009.\label{sfxt6:tab:bat_triggers} }
 \begin{tabular}{rllllrllll}
 \hline
 \noalign{\smallskip}
MJD &Date &Time$^{\mathrm{a}}$ &BAT Trigger N.$^{\mathrm{b}}$ &S/N$^{\mathrm{c}}$   \\
  \noalign{\smallskip}
 \hline
 \noalign{\smallskip}							        	  
 & &IGR~J16479$-$4514  \\
\noalign{\smallskip\smallskip}
54860	&2009-01-29 &06:32:06--06:48:1 &341452 (NFI) &10.68 \\
54895	&2009-03-04 & 17:43:07--18:13:47 \\ 
54915	&2009-03-25 & 00:13:07	         \\ 
54958	&2009-05-06 & 19:06:51	        \\
54986	&2009-06-03 & 23:45:07	          \\
55079	&2009-09-05 & 00:09:35	          \\
55116	&2009-10-12 & 03:25:31--03:47:39  \\
\noalign{\smallskip\smallskip}							        	  
 & &XTE~J1739$-$302 &   \\						        	  
\noalign{\smallskip\smallskip}
54901 &	2009-03-10 &18:18:35--18:45:15 &346069  (NFI)$^{\mathrm{d}}$ &6.81 \\
55079 &	2009-09-05 &05:00:27\\
55107 &	2009-10-03 &01:07:15-01:16:51 \\
\noalign{\smallskip\smallskip}							        	  
 & &IGR~J17544$-$2619 &   \\						        	  
\noalign{\smallskip\smallskip}
 54945  & 2009-04-24  & 11:19:15 &   \\   
 54988  & 2009-06-06  & 07:49:00  &354221 (NFI)$^{\mathrm{e}}$  &8.15   \\   
 55002  & 2009-06-20  & 02:45:39 &   \\  
 55024  & 2009-07-12  & 10:59:55 &   \\  
 55063  & 2009-08-20  & 03:16:43 &  \\ 
 55074  & 2009-08-30  & 12:33:55 &   \\ 
 55132  & 2009-10-28  & 19:24:11 &  \\   
  \noalign{\smallskip}
  \hline
  \end{tabular}
  \end{center}
  \begin{list}{}{}
  \item[$^{\mathrm{a}}$]{Time of the start of the BAT trigger, or the time range when on-board detections were obtained. } 
  \item[$^{\mathrm{b}}$]{BAT regular trigger, as was disseminated through GCNs. NFI indicates that there are data from the narrow-field instrument. 
} 
  \item[$^{\mathrm{c}}$]{On-board image significance in units of $\sigma$.}
  \item[$^{\mathrm{d}}$]{This work, Sect~\ref{sfxt6:spec_out17391}. }
  \item[$^{\mathrm{e}}$]{This work, Sect~\ref{sfxt6:spec_out17544}. }
  \end{list}  
\end{table}

The 15--50\,keV BAT Transient Monitor light curves collected from 2007 October 26 to  2009 November 3,
are shown in Fig.~\ref{sfxt6:fig:batlcvs}, rebinned at a 4\,d resolution.
The two gaps in each of the BAT light curves preceded and followed by points with large error bars 
are an artifact of \sw's Sun-avoidance pointing strategy.   
When the sources are near the Sun, they will be found only at the edges of the BAT FOV, 
where the sensitivity is reduced and for a few weeks each year the sources cannot be observed 
at all. 

In Table~\ref{sfxt6:tab:bat_triggers} we report the BAT on-board detections in the 15--50\,keV band. 
If an alert was generated, a BAT trigger was assigned (Column 4) and notices were 
disseminated\footnote{http://gcn.gsfc.nasa.gov/gcn/swift\_grbs.html}.  
For some of these triggers a burst response (slew and repointing of the NFI) was also initiated,
depending on GRB observing load, observing constraints, and interest in the sources.
More details on several of these triggers can be found in the papers of our series
(see Table~\ref{sfxt6:tab:campaign}, for references). 
The combination of all these data show how active the sources are outside the bright outbursts
when observed by the BAT.

\begin{figure}
\begin{center}
\vspace{-3truecm}
\centerline{\includegraphics[width=9cm,angle=0]{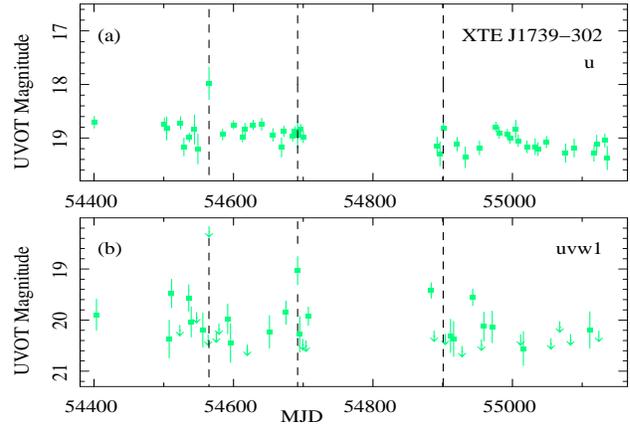}}
\vspace{-4truecm}
\caption{UVOT light curves of  XTE~J1739--302. The filters used are 
indicated in each panel. The vertical dashed lines mark the BAT outbursts. 
   }
	\label{sfxt6:fig:uvot_lcv_17391}   
       \end{center}
       \end{figure}

\begin{figure}
\begin{center}
\centerline{\includegraphics[width=3.3cm,angle=270]{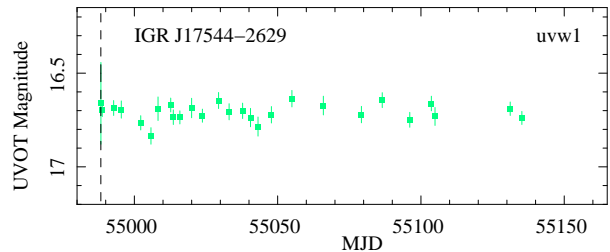}}
\caption{UVOT $uvw1$ light curve of IGR~J17544$-$2619. 
The vertical dashed lines mark the BAT outbursts. 
}
	\label{sfxt6:fig:uvot_lcv_17544}   
       \end{center}
       \end{figure}

       \subsection{UVOT light curves}

The UVOT performed observations simultaneously with the XRT, throughout most
of the \sw/XRT monitoring of the SFXTs.  
Figure~\ref{sfxt6:fig:uvot_lcv_17391}a,b shows the UVOT $u$ and $uvw1$ 
light curves of XTE~J1739$-$302 of the whole campaign. 
The dashed vertical lines mark the X--ray outbursts. 
The $u$ and $uvw1$ magnitudes show variability of marginal statistical significance;  
a fit against a constant yields  
$\chi^2_{\nu}(u)=2.13$ for 45 degrees of freedom (dof), 
null hypothesis probability nhp$=1.5\times10^{-5}$ ($\sim 4\,\sigma$), 
and $\chi^2_{\nu}(uvw1)=2.18$ for 20 dof, nhp$=1.6\times10^{-3}$ ($\sim 3\,\sigma$).
The $uvw2$ magnitudes are mostly upper limits, and the 5 detections yield a mean of
$uvw2=20.42\pm0.14$ mag. 
In \citet{Romano2009:sfxts_paperV} we reported a marginally significant (2--3 $\sigma$) 
increase of the $u$ magnitude during the first recorded outburst of this source (2008-04-08, MJD~54565), 
while during the second outburst (2008-08-13, MJD~54692) the $u$ magnitude is consistent with the 
mean for the whole campaign. The third reported outburst (2009-03-10, MJD~54901), 
like the second, shows no variation of the $u$ magnitude, with respect to the mean 
of the campaign. 
Observations were also taken once in the $v$ and $b$ filters during 2009 
as a follow-up ToO (obs.\ 00030987132),  
where we measure $v=15.25\pm0.03$ and $b=18.21\pm0.07$ mag. Both values are 
consistent with the ones obtained during and away from outbursts.

In Fig.~\ref{sfxt6:fig:uvot_lcv_17544} we show the UVOT $uvw1$ 
light curve of IGR~J17544$-$2619.  It is remarkably stable, as  
a fit against a constant yields  
$\chi^2_{\nu}(uvw1)=1.08$ for 26 dof, nhp$=0.35$. 
There are also a few observations performed in the other filters. 
We obtain $uvm2=20.3\pm0.3$ mag, $uvm2=20.0\pm0.3$ mag, $uvm2=19.9\pm0.2$ mag, $uvm2=19.9\pm0.3$ mag 
(obs.\ 00035056089, 00035056141, 00035056142, 00035056148, respectively), 
$uvw2=18.16\pm0.06$  mag and $uvw2=18.05\pm0.06$ mag (obs.\ 00035056143 and 00035056144), and 
$u=15.25\pm0.09$ mag (obs.\ 00035056145). 
During the 2009 June 6 outburst (obs.\ 354221000, MJD~54988)  all filters were used.  
The observed magnitudes are $u=15.16\pm0.01$ mag, $b=14.55\pm0.04$ mag,  
$uvw1=16.7\pm0.2$ mag.
The $uvw1$ value is consistent with those obtained during the remainder of the campaign. 
The $b$ magnitude is very close to the coincidence limit, 
therefore was heavily corrected for coincidence loss.

\section{Outbursts in 2009    } \label{sfxt6:outbursts} 

The year 2009 opened with the outburst of \src16479 {} on January 29, 
which we reported on in \citet{Romano2009:sfxts_paperV}. 
Here we analyze the data from three more outbursts caught from the sources in our
sample.  
 
       \subsection{IGR~J17391$-$3021} \label{sfxt6:spec_out17391}

\begin{figure}
\begin{center}
\centerline{\includegraphics[width=9cm,height=5cm,angle=0]{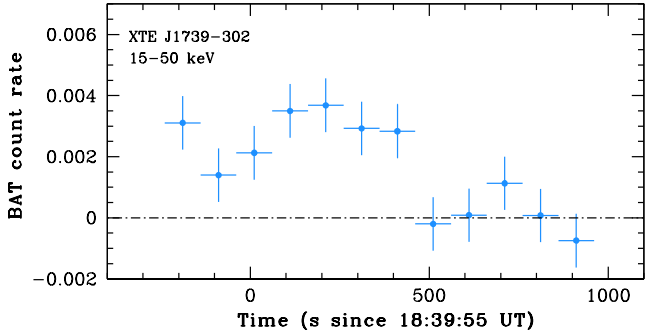}}
\caption{BAT light curve in the 15--50\,keV band of the 2009 March 10 outburst 
of XTE~J1739$-$302, in units of counts s$^{-1}$ det$^{-1}$. 
}
	\label{sfxt6:fig:bat_lcv_17391}   
       \end{center}
       \end{figure}

\xte17391 {} triggered the BAT on 2009 March 10 at 18:39:55 UT
(image trigger=346069). 
We report the BAT light curve in the 15--25\,keV band in Fig.~\ref{sfxt6:fig:bat_lcv_17391}. 
The time-averaged spectrum (T$+$0.0 to T$+$320.0\,s) is best fit by a simple power-law model
with a photon index of $2.8_{-0.6}^{+0.8}$, and the 15--50\,keV flux is 
$7.6\times10^{-10}$  erg cm$^{2}$ s$^{-1}$. 
\sw\ did not slew immediately in response to this trigger, so no NFI data were 
collected simultaneously with the BAT data. 
Observation 00030987107 (1.9\,ks net exposure, see Table~\ref{sfxt6:tab:alldata17391}) 
was obtained as a high-priority ToO observation, instead. 
The light curve obtained between T+5257\,s and T+10538\,s since the trigger 
shows a mean count rate of 0.4--1 counts s$^{-1}$. 
The mean XRT/PC spectrum can be fit with an absorbed powerlaw with a photon index of 
$\Gamma=1.1\pm0.4$  
and an absorbing column density of $N_{\rm H}=(3\pm1)\times10^{22}$ cm$^{-2}$ 
($\chi^2_{\nu}=1.1$ for 25 dof).  
The mean unabsorbed 2--10 keV flux is $9\times10^{-11}$ erg cm$^{2}$ s$^{-1}$, 
which translates into a luminosity of $7\times10^{34}$ erg s$^{-1}$ 
(assuming a distance of 2.7\,kpc, \citealt{Rahoui2008}). 
Further ToO observations were performed on 2009 March 11 (obs.\ 00030987108, 00030987109 in 
Table~\ref{sfxt6:tab:alldata17391}), when the XRT count rate was down to a few $10^{-2}$
counts s$^{-1}$. 
These spectral results are in general agreement with those obtained
during previous outbursts, 
both in terms of hard photon index
and in terms of a relatively lower column density (comparable with the out-of-outburst values).

       \subsection{IGR~J17544$-$2619} \label{sfxt6:spec_out17544}

\begin{figure}
\begin{center}
\vspace{-0.5truecm}
\centerline{\includegraphics[width=9cm,height=7cm,angle=0]{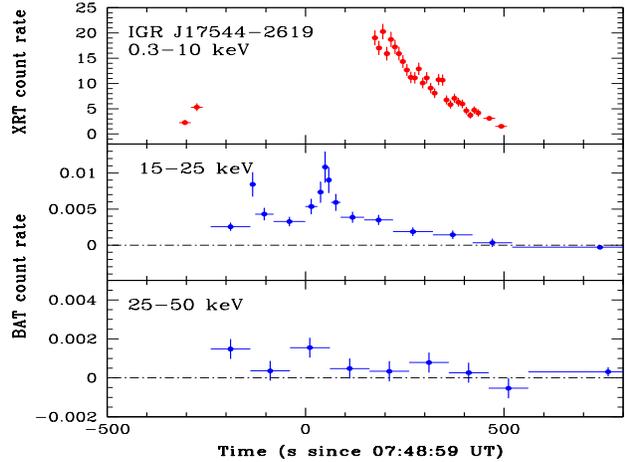}}
\caption{Light curves of the 2009 June 6 outburst of IGR~J17544$-$2619.
The XRT data preceding the outburst were collected 
as a pointed observation, part of our monitoring program. 
 The BAT data are in units of counts s$^{-1}$ det$^{-1}$.  
}
	\label{sfxt6:fig:xrt_bat_lcv_17544}   
       \end{center}
       \end{figure}

\begin{figure}
\begin{center}
\centerline{\includegraphics[height=9cm,angle=270]{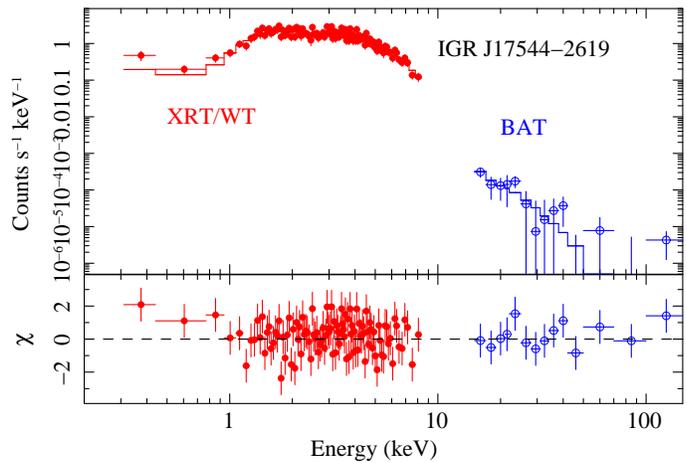}}
\caption{Spectroscopy of the 2009 June 6 outburst of IGR~J17544$-$2619. 
{\bf Top}: Data from the XRT/WT spectrum and simultaneous BAT spectrum 
fit with a {\sc cutoffpl} model. 
{\bf Bottom}: residuals in units of standard deviations.  
}
	\label{sfxt6:fig:xrt_bat_spec_17544}   
       \end{center}
       \end{figure}

 \begin{table}
 \begin{center}
 \caption{Spectral fits of simultaneous XRT and BAT data of IGR~J17544$-$2619 
during the 2009 June 06 outburst.  
}
 \label{sfxt6:tab:broadband_specfits}
 \begin{tabular}{llllllll}
 \hline
 \hline
 \noalign{\smallskip}
Model  &$N_{\rm H}$$^{\mathrm{a}}$ &$\Gamma$    &E$_{\rm c}$    &E$_{\rm f}$    &Flux$^{\mathrm{c}}$   &$\chi^{2}_{\nu}$/dof  \\
 \hline
\sc{POW}$^{\mathrm{b}}$  &$2.2_{-0.2}^{+0.3}$   &$1.7_{-0.1}^{+0.1}$   &       &    &  $2.6$      &$1.53/117$          \\
\sc{HCT}$^{\mathrm{b}}$  &$1.0_{-0.3}^{+0.2}$  &$0.6_{-0.4}^{+0.2}$  &$3_{-1}^{+1}$   &$8_{-3}^{+4}$   &  $1.9$      &$0.92/115$   \\
\sc{CPL}$^{\mathrm{b}}$ &$1.0_{-0.2}^{+0.3}$  &$0.4_{-0.3}^{+0.3}$  &$7_{-2}^{+4}$    &   &  $1.9$   &$0.94/116$         \\

\noalign{\smallskip}
  \hline
  \end{tabular}
  \end{center}
  \begin{list}{}{} 
 \item[$^{\mathrm{a}}$]{Absorbing column density is in units of $10^{22}$ cm$^{-2}$.}
   \item[$^{\mathrm{b}}$]{POW=simple absorbed powerlaw. 
HCT= absorbed powerlaw with high energy cutoff E$_{\rm c}$ (keV), e-folding energy E$_{\rm f}$ (keV). 
CPL=cutoff powerlaw with energy cutoff E$_{\rm c}$ (keV).}
  \item[$^{\mathrm{c}}$]{Unabsorbed 0.5--100\,keV flux is in units of $10^{-9}$ erg cm$^{-2}$ s$^{-1}$.}
  \end{list}
  \end{table}

\begin{figure*}
\begin{center}
\vspace{-2truecm}
\centerline{\includegraphics[width=18cm,height=20cm,angle=0]{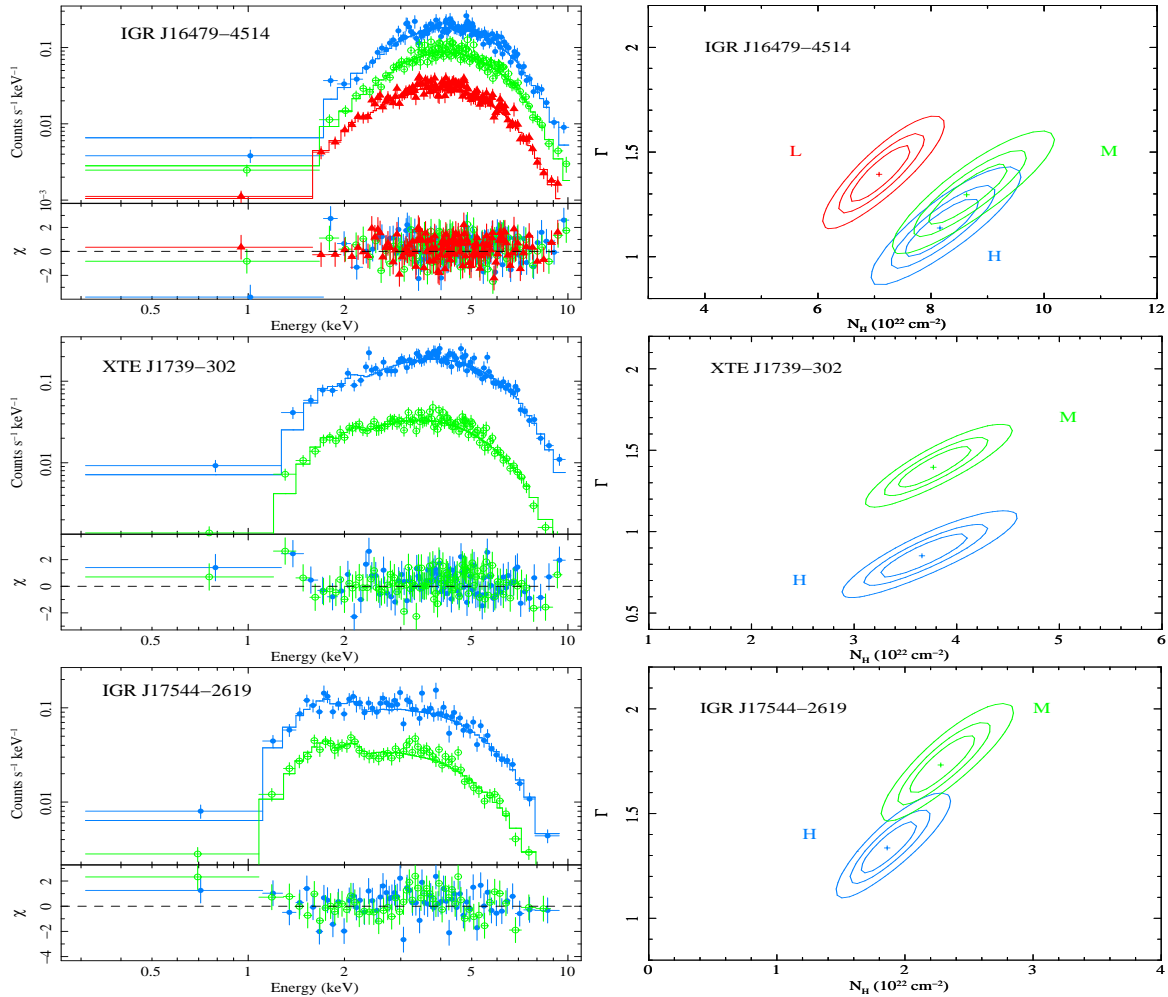}}
\vspace{-3.5truecm}
\caption[XRT spectroscopy, spectra and residuals]{Spectroscopy of the 2007--2009 
                  observing campaign.
            {\bf Left.} Upper panels: XRT/PC data fit with an absorbed power law. 
            Lower panels: the residuals of the fit (in units of standard deviations). 
            Filled blue circles, green empty circles, and red filled triangles
            mark high, medium, and low states, respectively. 
            {\bf Right.} 
            The $\Delta \chi^2 = 2.3$, 4.61 and 9.21 contour levels for the column density in 
            units of $10^{22}$ cm$^{-2}$ vs. the photon index, with best fits indicated by crosses. 
            The labels L, M, and H mark low, medium, and high states, respectively. 
                 }
	\label{sfxt6:fig:xrtspecfits}   
       \end{center}
       \end{figure*}

Two outbursts of IGR~J17544$-$2619 were caught during 2009. 
The first was observed by the BAT Transient Monitor 
as a sequence of bright flares. In the \sw{} pointings starting at
2009-03-15 23:52:40 UT  (1024\,s exposure) and 
at 2009-03-16 01:27:36 UT (1152\,s) the source reached  
$0.026\pm0.005$ counts s$^{-1}$ cm$^{-2}$ (115 mCrab, 15--50\,keV band) 
and $0.032\pm0.005$ counts s$^{-1}$ cm$^{-2}$ (140 mCrab),
respectively. It then faded below detectability before
re-brightening twice more on 2009 March 16: 95 mCrab in a 1152\,s 
exposure beginning 06:16:40 UT 
and then to 85 mCrab in a 384\,s exposure beginning at 09:25:52. 
XRT observations were performed as part of our regular monitoring 
on 2009 March 14 (see Table~\ref{sfxt6:tab:alldata17544}),  
$\sim46$\,h before the BAT outburst.  
A 3$\sigma$ upper limit was obtained at $1.6\times10^{-2}$ counts s$^{-1}$.
A later observation, obtained as a ToO on 2009 March 16 
($\sim16$\,h after the outburst), yielded a 3$\sigma$ upper limit at 
$1.3\times10^{-2}$ counts s$^{-1}$.

IGR~J17544$-$2619  triggered the BAT on 2009 June 06 at 07:48:59 UT 
(image trigger=354221).  \sw{} immediately slewed to the target, 
so that the NFIs started observing about 164\,s after the trigger. 
Figure~\ref{sfxt6:fig:xrt_bat_lcv_17544} shows the XRT and BAT light curves,
where it must be noted that the first two XRT points, preceding the outburst, 
were collected as a pointed observation, part of our monitoring program. 
The initial XRT burst data (WT mode in observation 00354221000, see 
Table~\ref{sfxt6:tab:alldata17544}, 170--446\,s since the trigger) show
a decaying light curve with a count rate that started at about 20 counts s$^{-1}$.
The following PC data (448--538s) seamlessly continue the decaying trend
down to about 2 counts s$^{-1}$. 
The WT spectrum, extracted during the peak of the outburst (with a grade 0 selection
to mitigate residual calibration uncertainties at low energies)
results in a hard X--ray emission. When fit with an absorbed power law, we obtain a 
photon index of $1.05_{-0.14}^{+0.15}$ and a column density of  
$N_{\rm H}=(1.3\pm0.2)\times 10^{22}$ cm$^{-2}$ 
($\chi^2_{\nu}=1.04$ for 104 dof).    
The unabsorbed 2--10\,keV flux is $9.5\times10^{-10}$ erg cm$^{-2}$ s$^{-1}$.  
The PC spectrum 
(unabsorbed 2--10\,keV flux $\sim 2\times10^{-10}$ erg cm$^{-2}$ s$^{-1}$) 
fitted with a power-law model yields a photon index of $1.1\pm0.7$  and 
$N_{\rm H}=(1.2_{-0.8}^{+1.2})\times 10^{22}$  cm$^{-2}$ 
[\citet{Cash1979} statistics and spectra binned to 1 count bin$^{-1}$; 
C--stat$=69.08$ for 47\,\% of $10^{4}$ Monte Carlo 
realizations with fit statistic less than C--stat],  
consistent with the WT data fit.

BAT mask-weighted spectra were extracted over time 
intervals strictly simultaneous with XRT data. 
We fit the simultaneous BAT$+$XRT spectra in the time interval 
170--446\,s since the BAT trigger, in the 0.3--10\,keV and 15--150\,keV
energy bands for XRT and BAT, respectively. 
Factors were included in the fitting to allow for normalization 
uncertainties between the two instruments, constrained within their 
usual ranges (0.9--1.1). 
Table~\ref{sfxt6:tab:broadband_specfits} reports our fits.
A simple absorbed power-law model is clearly an inadequate representation 
of the broad band spectrum with a $\chi^2_{\nu}=1.53$ for 117 dof). 
We then considered other curved models typically used to describe the X--ray emission from 
accreting pulsars in HMXBs, such as an absorbed  power-law model with a 
high energy cut-off ({\sc highecut} in \textsc{xspec}) and an absorbed  power-law model with an
exponential cutoff  ({\sc cutoffpl}).  
The latter models provide a satisfactory deconvolution of the 0.3--150\,keV emission, 
resulting in a hard powerlaw-like spectrum below 10\,keV, 
with a roll over of the high energies when simultaneous XRT and BAT data 
fits are performed. Figure~\ref{sfxt6:fig:xrt_bat_spec_17544} shows the fits
for the {\sc cutoffpl} model.

\section{Out-of-outburst X--ray spectroscopy}  \label{sfxt6:spec_out_of_outburst}

\begin{table*}
 \begin{center}
 \caption{XRT spectroscopy of the three SFXTs  (2007--2009 data set).\label{sfxt6:tab:xrtspecfits} }
 \begin{tabular}{crrcccccclc}
 \hline
 \noalign{\smallskip}
Name  & Spectrum & Rate &$N_{\rm H}$& Parameter  &
            &Flux$^{a}$   & Luminosity$^{b}$ &$\chi^{2}_{\nu}$/dof$^{c}$  \\ 
Absorbed power law  & & (counts s$^{-1}$)& ($10^{22}$~cm$^{-2}$)    & $\Gamma$  &   &(2--10 keV)  & (2--10 keV)  &Cstat(\%)  \\
  \noalign{\smallskip}
 \hline
 \noalign{\smallskip}
  IGR~J16479$-$4514  & high     & $>$0.55      &$8.2_{-0.7}^{+0.8}$ &$1.1_{-0.2}^{+0.2}$ &      &120 &5 &$1.2/193$	      \cr
                     & medium   & [0.22--0.55[ &$8.6_{-0.8}^{+0.8}$ &$1.3_{-0.2}^{+0.2}$ &      &53  &2 &$0.9/197$	      \cr
                     & low      & [0.06--0.22[ &$7.1_{-0.6}^{+0.6}$ &$1.4_{-0.1}^{+0.2}$ &      &17  &0.7  &$1.0/205$	      \cr
                     & very low$^{d}$ & $<$0.06 &$3.3_{-0.0}^{+0.4}$  &$1.8_{-0.2}^{+0.3}$ &      &1.3  &0.04   &$302.5(99.5)$	      \cr
\noalign{\smallskip\hrule\smallskip}
     XTE~J1739$-$302 & high     &$>$0.405      &$3.7_{-0.4}^{+0.5}$ &$0.8_{-0.1}^{+0.2}$  &       &120 &1       &$1.0/160$	       \cr
                     & medium   &[0.07--0.405[ &$3.8_{-0.4}^{+0.4}$ &$1.4_{-0.1}^{+0.1}$  &       &18  &0.2   &$0.9/164$		\cr
                     & very low$^{d}$ &$<$0.07 &$1.7_{-0.0}^{+0.1}$  &$1.4_{-0.2}^{+0.2}$  &       &0.5  &0.004   &$321.9(98.6)$	       \cr
\noalign{\smallskip\hrule\smallskip}							      
   IGR~J17544$-$2619 & high     &$>$0.25       &$1.9_{-0.2}^{+0.3}$ &$1.3_{-0.1}^{+0.1}$ &      &46 &0.8  &$1.0/118$	         \cr
                     & medium   &[0.07--0.25[  &$2.3_{-0.3}^{+0.3}$ &$1.7_{-0.2}^{+0.2}$ &      &14 &0.3   &$1.0/108$ 	 \cr
                     & very low$^{d}$ &$<$0.07 &$1.1_{-0.0}^{+0.1}$  &$2.1_{-0.2}^{+0.2}$ &      &0.2 &0.003   &$183.1(85.1)$	 \cr 
  \noalign{\smallskip}
 \hline
  \end{tabular}
  \end{center}
  \begin{list}{}{}
  \item[$^{\mathrm{a}}$]{Average observed 2--10\,keV fluxes in units of 10$^{-12}$~erg~cm$^{-2}$~s$^{-1}$.}
  \item[$^{\mathrm{b}}$]{Average 2--10\,keV X--ray luminosities in units of 10$^{35}$~erg~s$^{-1}$ calculated 
         adopting distances determined by \citet{Rahoui2008}. }
  \item[$^{\mathrm{c}}$]{Reduced $\chi^{2}$ and dof, or Cash statistics C--stat and percentage of realizations  ($10^4$ trials) with statistic $>$ Cstat. }
  \item[$^{\mathrm{d}}$]{Fit performed with constrained column density (see Sect~\ref{sfxt6:spec_out_of_outburst}). }
  \end{list}  
\end{table*}

We characterize the spectral properties of the sources in several states, 
by accumulating the events in each observation when 
the source was not in outburst and a detection was achieved. 
We selected three count rate levels, $CR_{1}$, $CR_{2}$, and $CR_{3}$ 
(reported in Table~\ref{sfxt6:tab:xrtspecfits}) 
that would yield comparable statistics 
in the ranges $CR_{1}<CR<CR_{2}$ (low), $CR_{2}<CR<CR_{3}$ (medium), and $CR>CR_3$ (high). 
If the statistics did not allow this, then we only considered two intensity levels 
(high and medium). 
Exposure maps and ARF files were created following the procedure described in 
\citet{Romano2009:sfxts_paperV}. 
The spectra were rebinned with a minimum of
20 counts per energy bin to allow $\chi^2$ fitting. 
Each spectrum was fit in the energy range 0.3--10\,keV 
with a single absorbed power law.  

We accumulated all data for which no detections were obtained as single exposures 
(whose combined exposure is $\Delta T_{\Sigma}$, reported in Table~\ref{sfxt6:tab:dutycycle}, column 2), 
and performed a detection. The resulting cumulative mean count rate for each object is reported in 
Table~\ref{sfxt6:tab:dutycycle} (column 5). 
Spectra were also extracted from these event lists. 
They consisted of $\sim 200$--$300$ counts each, 
so Cash statistics and spectra binned to 1 count bin$^{-1}$ 
were used, instead. 
When fitting with free parameters, the best fit value for $N_{\rm H}$ turned out to be 
consistent with 0, i.e., well below the column derived from optical spectroscopy. 
We therefore performed fits by adopting as lower limit on the absorbing 
column the value derived from the Galactic extinction estimate along the line of 
sight to each source from \citet{Rahoui2008}, with a conversion into 
Hydrogen column, $N_{\rm H} = 1.79 \times 10^{21} A_V $ cm$^{œôø²2}$ \citep{avnh}. 
Fig.~\ref{sfxt6:fig:xrtspecfits} shows the spectra and contour plots of photon 
index vs.\ column density; the spectral parameters are reported in 
Table~\ref{sfxt6:tab:xrtspecfits}, where 
we also report the average 2--10\,keV luminosities 
calculated by adopting distances of 4.9~kpc for IGR~J16479$-$4514, 2.7~kpc 
for XTE~J1739$-$302, and 3.6~kpc for IGR~J17544$-$2619 
determined by \citet{Rahoui2008} 
from optical spectroscopy of the supergiant companions. 
We also performed fits with an absorbed black body, obtaining consistent 
results with \citet[][]{Romano2009:sfxts_paperV}.

\section{Discussion\label{sfxt6:discussion}}

In this paper we report the results of a monitoring campaign with \sw\ 
that spans more than a two-year baseline.
Thanks to the unique characteristics of \sw\  we can  
investigate the properties of SFTXs on 
several timescales (from minutes to days, to years)
and in several intensity states (bright flares,
intermediate intensity states,  and down to almost quiescence).

\begin{figure}
\begin{center}
\centerline{\includegraphics[width=9cm]{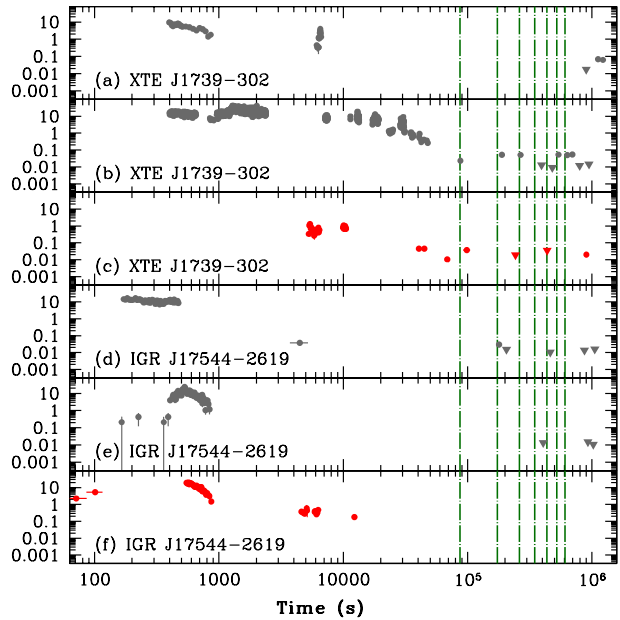}}
\caption[XRT light curves]{Light curves of the outbursts of SFXTs followed by {\it Swift}/XRT
referred to their respective BAT triggers. Points denote detections, triangles 3$\sigma$ upper limits.
Red data points refer to observations presented here for the first time, 
while grey data points refer to data presented elsewhere.  
Note that where no data are plotted, no data were collected.  
Vertical dashed lines mark time intervals equal to 1 day, up to a week. 
Panels  (a), (b) show the flares from XTE~J1739$-$302
\citep[2008-04-08, 2008-08-13; ][]{Sidoli2009:sfxts_paperIII,Sidoli2009:sfxts_paperIV}, while 
panel (c) refers to the 2009 March 10 flare reported here for the first time. 
Panels  (d), (e) show the flares from IGR~J17544--2916 
\citep[2008-03-31, 2008-09-04; ][]{Sidoli2009:sfxts_paperIII,Sidoli2009:sfxts_paperIV}, while 
panel (f) refers to the 2009 June 06 flare also 
reported here for the first time.  
  }
		\label{sfxt6:fig:sfxtlightcurves} 
        \end{center}
        \end{figure}

During the second year of monitorning 2 different sources flared
3 times, and in 2 cases we obtained multi-wavelength 
observations (see Table~\ref{sfxt6:tab:campaign}).  
\xte17391 {} triggered the BAT on 2009 March 10. 
Our results are in general agreement with those obtained
during previous outbursts,  
both in 
terms of hard photon index and in terms of a relatively lower column density 
(comparable with the out-of outburst values). 
When IGR~J17544$-$2619 
triggered the BAT on 2009 June 06, simultaneous BAT and XRT 
data were collected, thus allowing broad band spectroscopy. 
The soft X--ray spectral properties observed during this flare are 
generally consistent with those observed with $Chandra$ 
\citep[$\Gamma=0.73\pm{0.13}$, $N_{\rm H}=(1.36\pm{0.22}$)$\times$10$^{22}$~cm$^{-2}$, 
peak flux of $\sim$3$\times$10$^{-9}$~erg~cm$^{-2}$~s$^{-1}$; ][]{zand2005}.
However, when XRT and BAT data are jointly fit, an 
absorbed power-law model is inadequate in fitting the broad band spectrum,
and more curvy models are required. We considered 
an absorbed power-law model with a high energy cut-off 
and an absorbed  power-law model with an exponential cutoff,
models typically used to describe the X--ray emission from 
accreting neutron stars in HMXBs.  
We obtained a good deconvolution of the 0.3--150\,keV spectrum, characterized by 
a hard power law below 10\,keV and a well-constrained cutoff at higher energies.

In Fig.~\ref{sfxt6:fig:sfxtlightcurves} we compare the light curves 
of the most recent outbursts of \xte17391 {} and IGR~J17544$-$2619 (red points)
observed by \sw{} during the second year of monitoring, together 
with the previous outbursts of these sources followed by \sw/XRT (grey points). 
The light curves are referred to their respective BAT triggers
(unless the flare did not trigger the BAT, Fig.~\ref{sfxt6:fig:sfxtlightcurves}e). 
The most complete and deep set of X--ray observations of an outburst of a SFXT is 
the one of the periodic SFXT IGR~J11215$-$5952 
\citep[][]{Romano2007,Sidoli2007,Romano2009:11215_2008}, 
which was instrumental in discovering that the accretion phase during the 
bright outbursts lasts much longer than a few hours, as seen by lower-sensitivity 
instruments, and contrary to what
initially thought at the time of the discovery of this new class 
of sources \citep[e.g., ][]{Sguera2005}.
This behaviour is now found to be a common characteristic of the whole sample
of SFXTs followed by \sw{}, as our observations on \src16479 , IGR~J08408$-$4503, 
and SAX~J1818.6$-$1703 demonstrate \citep[][respectively; the vertical dashed lines 
in Fig.~\ref{sfxt6:fig:sfxtlightcurves} mark time intervals equal to 1 day, 
up to a week]{Romano2009:sfxts_paperV,Romano2009:sfxts_paper08408,Sidoli2009:sfxts_sax1818}.
In the same manner, our observations show that both the large dynamical range 
(up to 4 orders of magnitude) in flux
and a multiple-peaked structure of the light curves of the bright outbursts 
are equally common characteristics of the sample.  

In the optical/UV we see only marginal variability in the $u$ and $uvw1$ magnitudes 
of XTE~J1739$-$302, and  we could exclude a variation of flux corresponding to
the X--ray outburst. The $uvw1$ light curve of IGR~J17544$-$2619 is  
observed to be remarkably stable. This is consistent with the  optical/UV 
emission being  dominated by the constant contribution 
of the companion stars. 

\begin{figure*}
\begin{center}
\vspace{-1.5truecm}
\centerline{\includegraphics[width=20cm,height=34cm,angle=0]{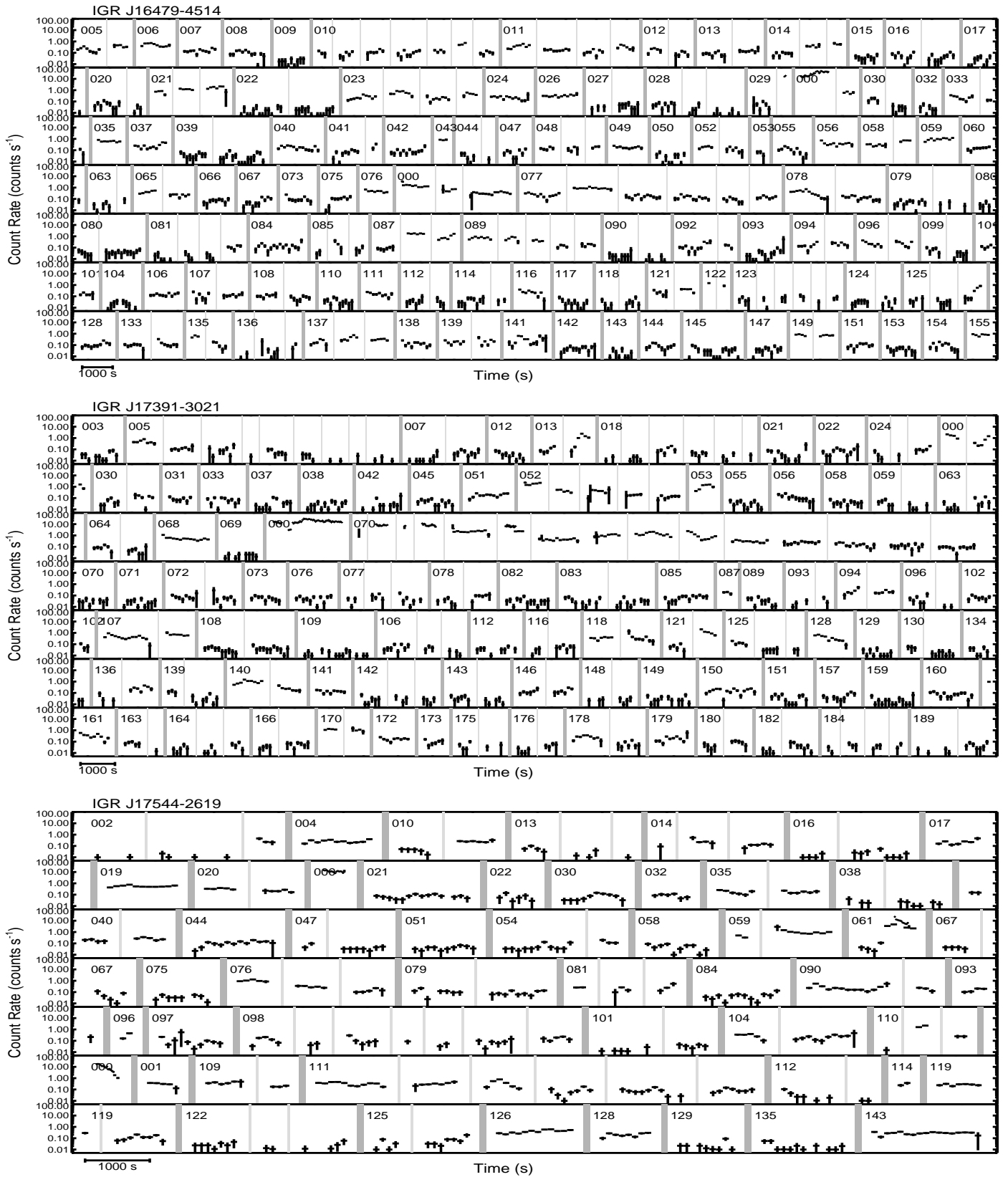}}
\vspace{-10.5truecm}
\caption[XRT light curves]{\sw/XRT (0.2--10\,keV) montage of time sequences. 
The numbers in the plot identify each observing sequence (see Table~\ref{sfxt6:tab:alldata16479}, 
\ref{sfxt6:tab:alldata17391}, and \ref{sfxt6:tab:alldata17544}), with 000 marking outburts.
Non observing intervals and orbital gaps have been cut out from the time axis and replaced 
by thick vertical bars to separate different sequences and thin grey bars to separate different 
orbits within each sequence.
Each point represents a 100\,s bin in PC and a 20\,s bin in WT modes. 
As a reference, on the bottom-left we show the 1000\,s time unit. 
The light curves are corrected for pile-up  PSF losses, vignetting and background-subtracted.  
                }
		\label{sfxt6:fig:montages} 
        \end{center}
        \end{figure*}

Variability in the X--ray flux is observed at all timescales and 
intensity ranges we can probe.  
Figure~\ref{sfxt6:fig:montages} shows, for each of the monitored sources, 
a montage of time sequences of the \sw/XRT data where detection was achieved 
(binned at a 100\,s and 20\,s resolution for PC and WT data, respectively) 
and with observing and orbital gaps removed from the time axis. 
Superimposed on the day-to-day variability (best shown in 
Fig.~\ref{sfxt6:fig:xrtlcvs} which is binned to a day 
resolution) is intra-day flaring. 
As Fig.~\ref{sfxt6:fig:montages} shows, the latter involves flux 
variations up to one order of magnitude that can occur down to timescales 
as short as an XRT snapshot ($\la 1$\,ks).  
This remarkable short time scale variability cannot be accounted for 
by accretion from a homogeneous wind. On the contrary, it can be naturally explained 
by the accretion of single clumps composing the donor wind, 
independently on the detailed geometrical and kinematical properties of the 
wind and the properties of the accreting compact object. 
If, for example, we assume that each of these short flares is caused 
by the accretion of a single clump onto the NS \citep[e.g., ][]{zand2005},
then its mass can be estimated \citep[][]{Walter2007} as 
$M_{\rm cl}= 7.5\times 10^{21} \,\, (L_{\rm X, 36}) (t_{\rm fl, 3{\rm ks}})^{3}$ g,
where $L_{\rm X, 36}$ is the average X-ray luminosity in units of $10^{36}$ erg s$^{-1}$,
$t_{\rm fl, 3{\rm ks}}$ is the duration of the flares in units of 3\,ks. 
We can confidently identify flares down to a count rate in the order of 
0.1\,counts s$^{-1}$ (within a snapshot of about 1\,ks; see Fig.~\ref{sfxt6:fig:montages}); 
these correspond to luminosities in the order of 2--$6\times10^{34}$ erg s$^{-1}$, which yield
 $M_{\rm cl} \sim 0.3$--$2\times10^{19}$ g. 
These masses are about those expected \citep[][]{Walter2007} to be
responsible of short flares, below the \inte{}  detection threshold and which,
if frequent enough, may significantly contribute to the mass-loss rate.

Our \sw\  monitoring campaign has demonstrated for the first time that 
X--ray emission from SFXTs is present outside the bright outbursts
\citep[][]{Sidoli2008:sfxts_paperI,Romano2009:sfxts_paperV}, 
although at a much lower level 
thus showing how frequent and typical in SFXTs is accretion at a low level rate.
In this paper we refined the intensity 
selected spectroscopy of the out-of-outburst emission,
by adopting absorbed power laws 
which yield hard power law photon indices ($\Gamma$$\sim$1--2). 
All these results are consistent with our findings in 
\citet[][]{Romano2009:sfxts_paperV} 
and show that accretion occurs over several orders of magnitude 
in luminosity (3 for \xte17391 {} and 2 for the others; Table~\ref{sfxt6:tab:xrtspecfits}),
even when excluding the bright outbursts. 
In particular, the lowest luminosities we could study with \sw\ 
are 4$\times$10$^{32}$~erg~s$^{-1}$ and 3$\times$10$^{32}$~erg~s$^{-1}$ 
(2--10 keV; `very low' intensity level in Table~\ref{sfxt6:tab:xrtspecfits}) 
for XTE~J1739--302 and IGR~J17544$-$2619, respectively. 
These low luminosities are not consistent with Bondi-Hoyle accretion from a 
spherically symmetric steady wind \citep{BondiHoyle44}, 
as previously reported by \citet{Drave2010:17391_3021_period} and 
\citet{Clark2009:17544-2619period} for these two SFXTs, thus arguing for 
a clumpy nature of the accreting wind. 
This is consistent with our findings on short timescale variability, which 
also requires inhomogeneities in the wind.

\begin{figure}
\begin{center}
\centerline{\includegraphics[width=5.cm,height=9.5cm,angle=90]{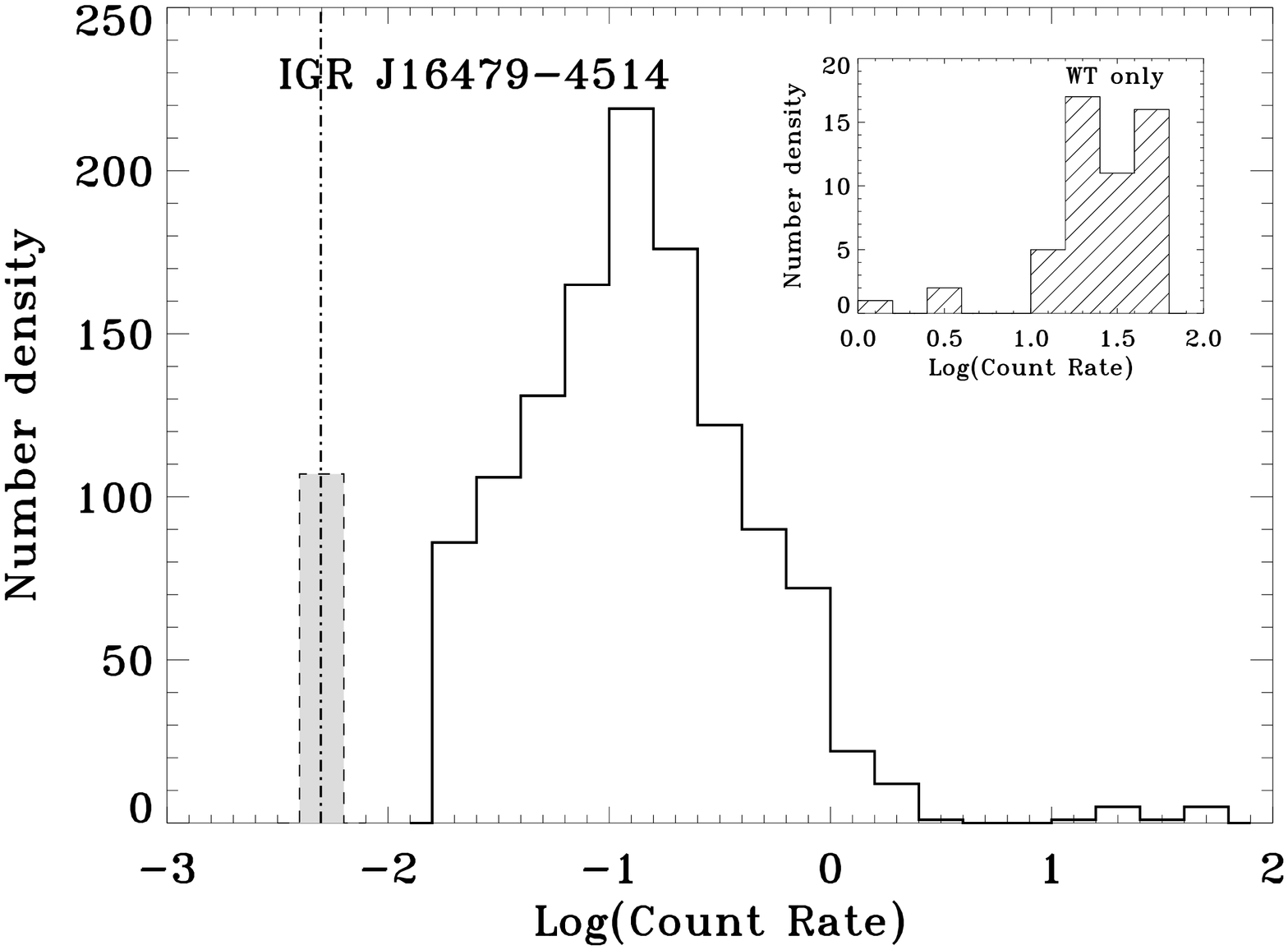}}
\centerline{\includegraphics[width=5.cm,height=9.5cm,angle=90]{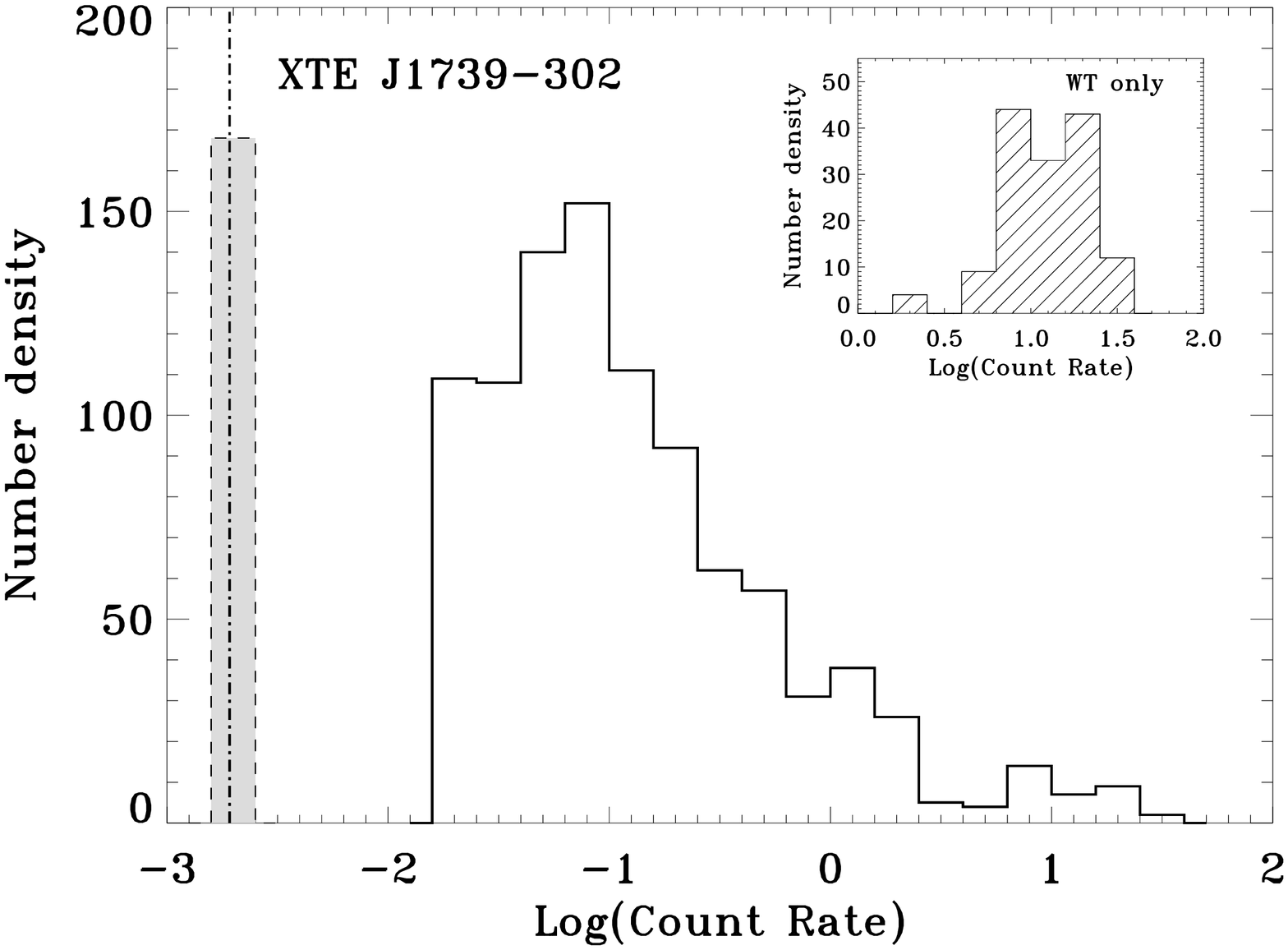}}
\centerline{\includegraphics[width=5.cm,height=9.5cm,angle=90]{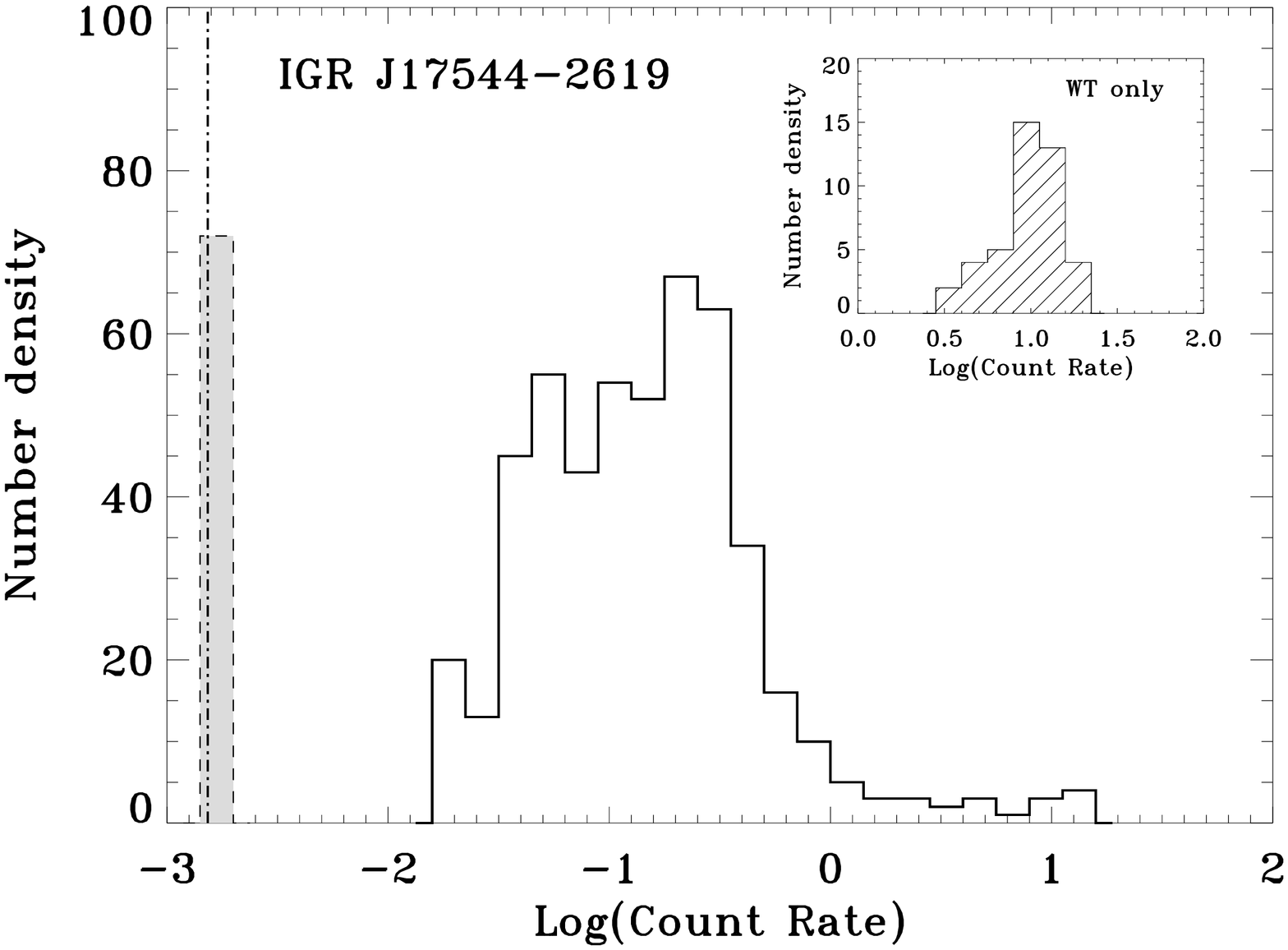}}
\caption[XRT light curves]{Distribution of the count rates when the XRT light curves are binned at 100\,s.  
The vertical lines correspond to the background. The hashed histograms are points which are consistent with 
a zero count rate. The insets show the subset of WT data only, when binned at 20\,s. }
		\label{sfxt6:fig:histos} 
        \end{center}
        \end{figure}

In the following, we assess the percentage of time each source spends at a given flux state.  
Figure~\ref{sfxt6:fig:histos} shows the distributions of the observed count rates 
after removal of the observations where a detection was not achieved 
 (same sample as in Fig.~\ref{sfxt6:fig:montages}, with PC and WT data binned  at 100\,s).  
In all cases a roughly Gaussian shape is observed, 
with a broad peak at $\approx 0.1$ counts s$^{-1}$, 
and a clear cut at the detection limit for 100\,s at the low end.
In particular, when the distributions are fit with a Gaussian function 
we find that their means and $\sigma$ are 
0.12 and 3.4 counts s$^{-1}$  for IGR~J16479$-$4514, 
0.06 and 4.9 counts s$^{-1}$  for XTE~J1739--302, and 
0.13 and 3.1 counts s$^{-1}$ for IGR~J17544$-$2619, respectively. 
This indicates that the most probable flux level at which a random observation
will find these sources, when detected, is 
$3 \times 10^{-11}$, $9\times 10^{-12}$, and $1 \times 10^{-11}$ erg cm$^{-2}$ s$^{-1}$ (unabsorbed 2--10\,keV,
obtained using the medium spectra in Table~\ref{sfxt6:tab:xrtspecfits}),  
corresponding to luminosities of $\sim 8\times 10^{34}$, $8\times 10^{33}$, and
 $2\times 10^{34}$ erg s$^{-1}$, respectively. 
Figure~\ref{sfxt6:fig:histos} also shows the hint of an excess at $\ga 10$ counts s$^{-1}$. 
Since this count rate is where the WT data of the outbursts are concentrated,
we also show them in the insets (dashed histograms)
binned at 20\,s resolution, which is more appropriate for such count rates. 
In this case, a clear peak emerges to represent the flaring state of the sources.

We have also calculated the time each source spends {\it undetected} down to a flux limit of 
1--3$\times10^{-12}$ erg cm$^{-2}$ s$^{-1}$, and we obtain ${\rm IDC} = 19, 39, 55$\,\%, 
for IGR~J16479$-$4514, XTE~J1739--302, and IGR~J17544$-$2619, respectively, 
with an estimated error of $\sim 5\,\%$. 
We note that these values are based on the whole 2-year campaign,
and that the IDC calculated with the observations of 2009 only is $21, 39, 55$\,\%.
These results can be compared with the ones reported in \citet{Romano2009:sfxts_paperV},
where we obtained ${\rm IDC} = 17, 39, 55$\,\%, for IGR~J16479$-$4514, XTE~J1739--302, and 
IGR~J17544$-$2619, respectively. 
We now have a baseline twice as long and almost a factor of two higher exposure;  
the sampling of the light curves has slightly changed during 2009.
Nevertheless, the calculated IDC values are remarkably stable, which  
is indicative of the robustness of this method with respect to the 
sampling pace.  
Considering our monitoring as a casual sampling of the light curve 
for two years, we can also infer that the time these sources spend 
in bright outbursts is between 3 and 5\,\% of the total.

\section*{Acknowledgments}

PR dedicates this effort to her grandmother, G.\ Ghedin; she could not follow 
her aspiration, yet worked hard so that others could. 
We acknowledge the input from our colleagues along the way during 
this large project, in particular 
A.\ Beardmore, M.M.\ Chester, L.\ Ducci, C.\ Guidorzi, T.\ Mineo,  M.\ Perri. 
We thank the {\it Swift} team duty scientists and science planners.   
We also thank the remainder of the {\it Swift} XRT, BAT, and UVOT teams,
S.\ Barthelmy and J.A.\ Nousek in particular, for their invaluable help 
and support with the planning and execution of the observing strategy. 
We also thank the anonymous referee for comments that helped improve the paper. 
This work was supported at PSU by NASA contract NAS5-00136. 
HAK was supported by the {\it Swift } project.

\bibliographystyle{aa}

\setcounter{table}{0} 
 \begin{table*}
 \begin{center}
 \caption{Observation log for IGR~J16479$-$4514. \label{sfxt6:tab:alldata16479}}
 \begin{tabular}{lllll}
 \hline
 \noalign{\smallskip}
 Sequence & Instrument/Mode & Start time (UT) &  End time (UT) &  Net Exposure \\
            &     & (yyyy-mm-dd hh:mm:ss) & (yyyy-mm-dd hh:mm:ss) &  (s) \\ 
  \noalign{\smallskip}
 \hline
 \noalign{\smallskip}
00030296096	&PC    &2009-02-19 21:23:39	&2009-02-19 23:11:56	 &	 1178	 \\
00030296097	&PC    &2009-02-23 15:15:37	&2009-02-23 17:08:58	 &	 1925	 \\
00030296098	&PC    &2009-02-26 20:38:22	&2009-02-26 22:21:58	 &	 902	 \\
00030296099	&PC    &2009-03-01 01:36:58	&2009-03-01 03:20:57	 &	 1014	 \\
00030296101	&PC    &2009-03-08 01:58:11	&2009-03-08 02:11:58	 &	 813	 \\
00030296102	&PC    &2009-03-12 12:00:26	&2009-03-12 12:13:56	 &	 794	 \\
00030296103	&PC    &2009-03-15 17:26:03	&2009-03-15 17:42:56	 &	 996	 \\
00030296104	&PC    &2009-03-19 22:41:22	&2009-03-19 22:57:50	 &	 987	 \\
00030296105	&PC    &2009-03-23 05:11:51	&2009-03-23 05:34:57	 &	 1348	 \\
00030296106	&PC    &2009-03-26 23:17:55	&2009-03-26 23:33:56	 &	 949	 \\
00030296107	&PC    &2009-03-29 20:20:10	&2009-03-29 22:05:55	 &	 1152	 \\
00030296108	&PC    &2009-04-02 10:51:38	&2009-04-02 12:44:57	 &	 1209	 \\
00030296110	&PC    &2009-04-09 22:57:46	&2009-04-09 23:13:56	 &	 962	 \\
00030296111	&PC    &2009-04-12 00:46:56	&2009-04-12 01:01:54	 &	 850	 \\
00030296112	&PC    &2009-04-16 01:03:12	&2009-04-16 02:49:58	 &	 871	 \\
00030296114	&PC    &2009-04-23 06:35:10	&2009-04-23 08:25:57	 &	 1128	 \\
00030296116	&PC    &2009-05-03 20:31:05	&2009-05-03 20:44:58	 &	 824	 \\
00030296117	&PC    &2009-05-07 07:45:46	&2009-05-07 08:02:56	 &	 1021	 \\
00030296118	&PC    &2009-05-14 21:31:15	&2009-05-14 23:13:56	 &	 771	 \\
00030296119	&PC    &2009-05-17 18:25:09	&2009-05-17 18:26:57	 &	 108	 \\
00030296120	&PC    &2009-05-21 02:33:56	&2009-05-21 04:15:57	 &	 817	 \\
00030296121	&PC    &2009-05-28 16:01:59	&2009-05-28 17:44:57	 &	 831	 \\
00030296122	&PC    &2009-06-04 09:00:40	&2009-06-04 10:22:56	 &	 66	 \\
00030296123	&PC    &2009-06-10 14:19:27	&2009-06-10 22:31:57	 &	 1087	 \\
00030296124	&PC    &2009-06-11 09:17:46	&2009-06-11 11:03:57	 &	 1166	 \\
00030296125	&PC    &2009-06-14 12:57:00	&2009-06-14 16:25:57	 &	 1344	 \\
00030296127	&PC    &2009-06-22 21:41:15	&2009-06-22 23:24:57	 &	 895	 \\
00030296128	&PC    &2009-06-25 15:35:42	&2009-06-25 15:51:58	 &	 960	 \\
00030296129	&PC    &2009-06-29 02:03:38	&2009-06-29 16:34:57	 &	 412	 \\
00030296130	&PC    &2009-07-02 22:56:56	&2009-07-02 23:15:57	 &	 1117	 \\
00030296131	&PC    &2009-07-06 23:03:27	&2009-07-06 23:41:57	 &	 929	 \\
00030296132	&PC    &2009-07-09 20:41:44	&2009-07-09 23:55:57	 &	 393	 \\
00030296133	&PC    &2009-07-12 07:45:29	&2009-07-12 09:29:58	 &	 1199	 \\
00030296134	&PC    &2009-07-19 18:04:32	&2009-07-19 18:13:56	 &	 564	 \\
00030296135	&PC    &2009-07-23 15:17:39	&2009-07-23 17:03:56	 &	 695	 \\
00030296136	&PC    &2009-07-26 13:36:28	&2009-07-26 17:00:57	 &	 974	 \\
00030296137	&PC    &2009-07-30 01:03:16	&2009-07-30 04:24:56	 &	 1608	 \\
00030296138	&PC    &2009-08-02 19:20:12	&2009-08-02 19:36:57	 &	 982	 \\
00030296139	&PC    &2009-08-06 12:58:49	&2009-08-06 14:46:57	 &	 1177	 \\
00030296141	&PC    &2009-08-13 02:28:41	&2009-08-13 02:48:45	 &	 817	\\ 
00030296142	&PC    &2009-08-20 17:46:23	&2009-08-20 18:04:57	 &	 1111	\\ 
00030296143	&PC    &2009-08-23 11:50:59	&2009-08-23 12:07:57	 &	 1004	 \\
00030296144	&PC    &2009-08-27 05:46:01	&2009-08-27 06:02:57	 &	 994	 \\
00030296145	&PC    &2009-08-30 05:52:17	&2009-08-30 06:20:57	 &	 1710	 \\
00030296146	&PC    &2009-09-10 15:02:38	&2009-09-10 15:18:58	 &	 967	 \\
00030296147	&PC    &2009-09-13 15:20:34	&2009-09-13 15:36:57	 &	 975	 \\
00030296148	&PC    &2009-09-17 02:48:04	&2009-09-17 03:05:57	 &	 1056	 \\
00030296149	&PC    &2009-09-20 06:00:19	&2009-09-20 07:42:56	 &	 785	 \\
00030296150	&PC    &2009-09-27 14:53:56	&2009-09-27 16:41:57	 &	 1127	 \\
00030296151	&PC    &2009-10-01 21:50:58	&2009-10-01 22:05:58	 &	 884	 \\
00030296152	&PC    &2009-10-04 02:43:42	&2009-10-04 02:59:57	 &	 943	 \\
00030296153	&PC    &2009-10-08 03:12:19	&2009-10-08 03:28:57	 &	 984	 \\
00030296154	&PC	&2009-10-18 13:39:21	&2009-10-18 13:54:56	&	905	\\
00030296155	&PC	&2009-10-25 08:04:27	&2009-10-25 08:13:56	&	569	\\
\hline
\noalign{\smallskip}
 \end{tabular}
  \end{center}
  \end{table*}

\setcounter{table}{1} 
 \begin{table*}
 \begin{center}
 \caption{Observation log for IGR~J17391$-$3021.\label{sfxt6:tab:alldata17391}}
 \begin{tabular}{lllll}
 \hline
 \noalign{\smallskip}
 Sequence & Instrument/Mode & Start time (UT) &  End time (UT) &  Net Exposure \\
            &     & (yyyy-mm-dd hh:mm:ss) & (yyyy-mm-dd hh:mm:ss) &  (s) \\ 
  \noalign{\smallskip}
 \hline
 \noalign{\smallskip}

00030987098	&PC	&2009-02-21 13:32:13	 &2009-02-21 15:19:57	  &	  1382    \\
00030987099	&PC	&2009-02-23 07:26:20	 &2009-02-23 09:11:58	  &	  1354    \\
00030987100	&PC	&2009-02-25 14:13:53	 &2009-02-25 15:56:58	  &	  1123    \\
00030987101	&PC	&2009-02-27 23:51:23	 &2009-02-27 23:59:44	  &	  487	  \\
00030987102	&PC	&2009-03-02 01:42:07	 &2009-03-02 01:57:57	  &	  939	  \\
00030987103	&PC	&2009-03-04 00:28:00	 &2009-03-04 03:43:57	  &	  811	  \\
00030987104	&PC	&2009-03-06 02:12:57	 &2009-03-06 05:35:56	  &	  833	  \\
00030987105	&PC	&2009-03-08 03:33:33	 &2009-03-08 03:53:56	  &	  1201    \\
00346069000     &BAT/evt &2009-03-10 18:36:00    &2009-03-10 18:56:02    &       1202    \\  
00030987107	&PC	&2009-03-10 20:07:35	 &2009-03-10 21:35:54	  &	  1883    \\
00030987108	&PC	&2009-03-11 05:52:47	 &2009-03-11 07:19:57	  &	  1975    \\
00030987109	&PC	&2009-03-11 13:28:45	 &2009-03-11 14:01:56	  &	  1976    \\
00030987106	&PC	&2009-03-11 20:16:24	 &2009-03-11 23:44:55	  &	  1361    \\
00030987110	&PC	&2009-03-13 13:55:19	 &2009-03-13 14:11:57	  &	  986	  \\
00030987111	&PC	&2009-03-15 19:04:51	 &2009-03-15 20:45:57	  &	  862	  \\
00030987112	&PC	&2009-03-21 05:03:14	 &2009-03-21 05:24:56	  &	  1288    \\
00030987113	&PC	&2009-03-23 22:53:06	 &2009-03-23 23:14:58	  &	  1257    \\
00030987114	&PC	&2009-03-25 16:54:16	 &2009-03-25 18:43:56	  &	  1322    \\
00030987115	&PC	&2009-03-27 17:10:56	 &2009-03-27 17:15:57	  &	  282	  \\
00030987116	&PC	&2009-03-30 17:19:23	 &2009-03-30 18:59:56	  &	  900	  \\
00030987117	&PC	&2009-04-01 14:14:04	 &2009-04-01 15:58:58	  &	  180	  \\
00030987118	&PC	&2009-04-06 16:19:14	 &2009-04-06 18:06:56	  &	  1353    \\
00030987120	&PC	&2009-04-11 07:15:28	 &2009-04-11 08:59:56	  &	  869	  \\
00030987121	&PC	&2009-04-12 05:47:30	 &2009-04-12 07:31:29	  &	  801	  \\
00030987122	&PC	&2009-04-17 14:04:05	 &2009-04-17 14:23:56	  &	  1186    \\
00030987124	&PC	&2009-04-22 01:44:10	 &2009-04-22 03:27:57	  &	  1155    \\
00030987125	&PC	&2009-04-24 01:58:43	 &2009-04-24 04:59:57	  &	  1189    \\
00030987126	&PC	&2009-04-29 17:03:39	 &2009-04-29 20:20:56	  &	  1162    \\
00030987127	&PC	&2009-05-01 20:17:41	 &2009-05-01 20:33:56	  &	  961	  \\
00030987128	&PC	&2009-05-04 12:15:55	 &2009-05-04 12:31:55	  &	  943	  \\
00030987129	&PC	&2009-05-06 23:45:39	 &2009-05-06 23:59:57	  &	  856	  \\
00030987130	&PC	&2009-05-08 00:03:19	 &2009-05-08 05:09:56	  &	  919	  \\
00030987131	&PC	&2009-05-15 17:00:32	 &2009-05-15 18:43:56	  &	  927	  \\
00030987132	&PC	&2009-05-18 15:18:35	 &2009-05-18 15:32:57	  &	  845	  \\
00030987133	&PC	&2009-05-20 01:18:12	 &2009-05-20 01:33:57	  &	  933	  \\
00030987134	&PC	&2009-05-22 23:55:04	 &2009-05-23 00:07:56	  &	  753	  \\
00030987135	&PC	&2009-05-25 09:40:45	 &2009-05-25 11:27:56	  &	  1102    \\
00030987136	&PC	&2009-05-27 06:23:01	 &2009-05-27 09:43:56	  &	  1295    \\
00030987137	&PC	&2009-05-29 21:36:23	 &2009-05-29 23:15:56	  &	  812	  \\
00030987139	&PC	&2009-06-03 04:22:19	 &2009-06-03 05:34:57	  &	  957	  \\
00030987140	&PC	&2009-06-10 14:26:00	 &2009-06-10 16:16:57	  &	  1484    \\
00030987141	&PC	&2009-06-12 04:41:14	 &2009-06-12 04:55:57	  &	  875	  \\
00030987142	&PC	&2009-06-14 13:04:58	 &2009-06-14 16:33:57	  &	  1465    \\
00030987143	&PC	&2009-06-17 11:59:25	 &2009-06-17 13:45:57	  &	  1253    \\
00030987144	&PC	&2009-06-19 13:34:34	 &2009-06-19 15:19:58	  &	  1106    \\
00030987145	&PC	&2009-06-22 21:50:01	 &2009-06-22 21:55:05	  &	  289	  \\
00030987146	&PC	&2009-06-24 20:34:06	 &2009-06-24 22:18:58	  &	  1009    \\
00030987147	&PC	&2009-06-26 04:38:26	 &2009-06-26 08:12:08	  &	  919	  \\
00030987148	&PC	&2009-06-29 18:07:36	 &2009-06-29 19:50:58	  &	  837	  \\
00030987149	&PC	&2009-07-01 13:15:25	 &2009-07-01 13:36:56	  &	  1288    \\
00030987150	&PC	&2009-07-03 17:58:50	 &2009-07-03 18:22:57	  &	  1439    \\
00030987151	&PC	&2009-07-08 18:46:26	 &2009-07-08 19:04:57	  &	  1086    \\
00030987152	&PC	&2009-07-10 17:34:35	 &2009-07-10 20:32:56	  &	  777	  \\
00030987154	&PC	&2009-07-17 09:58:00	 &2009-07-17 11:34:58	  &	  1049    \\
00030987155	&PC	&2009-07-20 13:27:01	 &2009-07-20 15:09:56	  &	  535	  \\
00030987156	&PC	&2009-07-22 18:22:50	 &2009-07-22 18:39:56	  &	  1019    \\
00030987157	&PC	&2009-07-24 15:24:43	 &2009-07-24 15:40:56	  &	  966	  \\
00030987159	&PC	&2009-07-29 10:44:26	 &2009-07-29 11:04:57	  &	  1218    \\
00030987160	&PC	&2009-08-03 21:05:26	 &2009-08-03 21:26:57	  &	  1273    \\
00030987161	&PC	&2009-08-05 21:10:28	 &2009-08-05 21:26:56	  &	  978	  \\
00030987162	&PC	&2009-08-07 08:19:05	 &2009-08-07 10:08:56	  &	  1262    \\
00030987163	&PC	&2009-08-10 21:30:24	 &2009-08-10 23:14:57	  &	  881	  \\
  \noalign{\smallskip}
 \hline
 \noalign{\smallskip}
  \end{tabular}
  \end{center}
  \end{table*}

\setcounter{table}{1}
 \begin{table*}
 \begin{center}
 \caption{Observation log for IGR~J17391$-$3021. Continued.}
 \begin{tabular}{lllll}
 \hline
 \noalign{\smallskip}
 Sequence & Instrument/Mode & Start time (UT) &  End time (UT) &  Net Exposure \\
            &     & (yyyy-mm-dd hh:mm:ss) & (yyyy-mm-dd hh:mm:ss) &  (s) \\ 
  \noalign{\smallskip}
 \hline
 \noalign{\smallskip}
00030987164	&PC	&2009-08-12 05:38:33	 &2009-08-12 08:57:57	  &	  1198    \\
00030987166	&PC	&2009-08-22 14:39:23	 &2009-08-22 16:56:56	  &	  900	  \\
00030987167	&PC	&2009-08-24 21:17:04	 &2009-08-24 21:27:56	  &	  648	  \\
00030987168	&PC	&2009-08-26 02:13:22	 &2009-08-26 02:31:58	  &	  1080    \\
00030987169	&PC	&2009-08-31 04:16:51	 &2009-08-31 04:42:56	  &	  1555    \\
00030987170	&PC	&2009-09-02 06:33:58	 &2009-09-02 08:15:57	  &	  570	  \\
00030987171	&PC	&2009-09-09 06:40:15	 &2009-09-09 08:23:56	  &	  1045    \\
00030987172	&PC	&2009-09-11 08:46:47	 &2009-09-11 09:01:57	  &	  904	  \\
00030987173	&PC	&2009-09-14 12:22:51	 &2009-09-14 12:31:57	  &	  528	  \\
00030987174	&PC	&2009-09-16 15:36:36	 &2009-09-16 15:52:57	  &	  968	  \\
00030987175	&PC	&2009-09-18 10:47:06	 &2009-09-18 12:35:57	  &	  834	  \\
00030987176	&PC	&2009-09-21 02:55:51	 &2009-09-21 04:39:56	  &	  808	  \\
00030987177	&PC	&2009-09-23 00:03:21	 &2009-09-23 12:59:58	  &	  844	  \\
00030987178	&PC	&2009-09-28 21:39:16	 &2009-09-28 23:24:56	  &	  1383    \\
00030987179	&PC	&2009-10-02 21:47:31	 &2009-10-02 22:02:57	  &	  904	  \\
00030987180	&PC	&2009-10-05 20:35:06	 &2009-10-05 22:19:57	  &	  1052    \\
00030987181	&PC	&2009-10-07 01:26:48	 &2009-10-07 03:11:56	  &	  1031    \\
00030987182	&PC	&2009-10-09 01:50:41	 &2009-10-09 03:33:55	  &	  1287    \\
00030987183     &PC    &2009-10-12 21:12:04	&2009-10-12 22:55:58	 &	 1426	 \\
00030987184     &PC    &2009-10-14 16:28:06	&2009-10-14 21:21:58	 &	 1080	 \\
00030987185     &PC    &2009-10-16 21:28:03	&2009-10-16 21:44:56	 &	 1008	 \\
00030987186     &PC    &2009-10-19 21:53:02	&2009-10-19 23:37:57	 &	 1065	 \\
00030987187     &PC    &2009-10-21 01:18:31	&2009-10-21 01:19:56	 &	 65	 \\
00030987188     &PC    &2009-10-26 22:39:02	&2009-10-26 22:44:56	 &	 333	 \\
00030987189     &PC    &2009-10-28 11:27:04	&2009-10-28 13:24:58	 &	 1663	 \\
00030987191     &PC    &2009-11-01 00:37:38	&2009-11-01 00:54:56	 &	 1030	 \\
  \noalign{\smallskip}
 \hline
 \noalign{\smallskip}
  \end{tabular}
  \end{center}
  \end{table*}

\setcounter{table}{2} 
\begin{table*}
 \begin{center}
 \caption{Observation log for IGR~J17544$-$2619.\label{sfxt6:tab:alldata17544}} 
 \begin{tabular}{lllll}
 \hline
 \noalign{\smallskip}
 Sequence & Instrument/Mode & Start time (UT) &  End time (UT) &  Net Exposure \\
            &     & (yyyy-mm-dd hh:mm:ss) & (yyyy-mm-dd hh:mm:ss) &  (s) \\ 
  \noalign{\smallskip}
 \hline
 \noalign{\smallskip}
00035056083	&PC	&2009-02-21 16:45:15	 &2009-02-21 18:32:57	  &	  1387    \\
00035056084	&PC	&2009-02-24 07:31:27	 &2009-02-24 07:50:58	  &	  1160    \\
00035056085	&PC	&2009-02-28 01:28:32	 &2009-02-28 01:45:57	  &	  1034    \\
00035056086	&PC	&2009-03-03 19:35:54	 &2009-03-03 21:18:57	  &	  784	  \\
00035056087	&PC	&2009-03-07 20:03:32	 &2009-03-07 21:41:56	  &	  268	  \\
00035056088	&PC	&2009-03-14 01:11:35	 &2009-03-14 01:28:56	  &	  1031    \\
00035056089	&PC	&2009-03-16 15:35:47	 &2009-03-16 17:17:37	  &	  2126    \\
00035056090	&PC	&2009-03-21 17:59:47	 &2009-03-21 19:40:56	  &	  1450    \\
00035056091	&PC	&2009-03-28 18:37:33	 &2009-03-28 18:53:56	  &	  946	  \\
00035056092	&PC	&2009-03-31 14:11:09	 &2009-03-31 14:15:57	  &	  281	  \\
00035056093	&PC	&2009-04-04 17:56:02	 &2009-04-04 18:02:56	  &	  410	  \\
00035056094	&PC	&2009-04-07 00:20:07	 &2009-04-07 00:39:57	  &	  916	  \\
00035056095	&PC	&2009-04-11 10:30:29	 &2009-04-11 12:14:58	  &	  889	  \\
00035056096	&PC	&2009-04-18 04:35:54	 &2009-04-18 04:38:57	  &	  163	  \\
00035056097	&PC	&2009-04-25 00:26:41	 &2009-04-25 00:42:58	  &	  698	  \\
00035056098	&PC	&2009-04-28 07:20:35	 &2009-04-28 13:42:55	  &	  2118    \\
00035056099	&PC	&2009-05-02 09:02:27	 &2009-05-02 09:18:57	  &	  968	  \\
00035056100	&PC	&2009-05-05 01:31:22	 &2009-05-05 03:12:57	  &	  962	  \\
00035056101	&PC	&2009-05-09 04:42:29	 &2009-05-09 06:27:56	  &	  1237    \\
00035056102	&PC	&2009-05-16 02:35:27	 &2009-05-16 04:25:57	  &	  865	  \\
00035056103	&PC	&2009-05-19 17:21:57	 &2009-05-19 19:03:57	  &	  879	  \\
00035056104	&PC	&2009-05-23 09:28:09	 &2009-05-23 11:16:57	  &	  1361    \\
00035056105	&PC	&2009-05-24 22:29:23	 &2009-05-25 00:15:56	  &	  959	  \\
00035056106	&PC	&2009-05-25 14:29:38	 &2009-05-26 06:27:56	  &	  565	  \\
00035056107	&PC	&2009-05-30 23:00:07	 &2009-05-30 23:08:56	  &	  519	  \\
00035056108	&PC	&2009-06-02 07:23:10	 &2009-06-02 09:02:56	  &	  400	  \\
00035056110	&PC	&2009-06-06 04:32:43	 &2009-06-06 09:08:57	  &	  332	  \\
00354221000	&BAT/evt&2009-06-06 07:45:04 	 &2009-06-06 08:05:06 	  &	  1202	  \\ 
00354221000	&WT	&2009-06-06 07:51:53	 &2009-06-06 07:56:29	  &	  276	  \\
00354221000	&PC	&2009-06-06 07:56:31	 &2009-06-06 07:58:01	  &	  90	  \\
00354221001	&PC	&2009-06-06 08:58:36	 &2009-06-06 09:05:57	  &	  417	  \\
00035056109	&PC	&2009-06-06 09:20:02	 &2009-06-06 11:09:57	  &	  776	  \\
00035056111	&PC	&2009-06-06 15:17:45	 &2009-06-06 20:48:54	  &	  4409    \\
00035056112	&PC	&2009-06-10 17:43:17	 &2009-06-10 19:30:57	  &	  1376    \\
00035056113	&PC	&2009-06-13 09:44:23	 &2009-06-13 11:29:58	  &	  1016    \\
00035056114	&PC	&2009-06-16 18:24:52	 &2009-06-16 18:27:57	  &	  143	  \\
00035056115	&PC	&2009-06-20 02:36:19	 &2009-06-20 07:39:57	  &	  1551    \\
00035056116	&PC	&2009-06-23 18:36:44	 &2009-06-23 18:55:59	  &	  1060    \\
00035056117	&PC	&2009-06-26 06:14:29	 &2009-06-26 06:27:09	  &	  745	  \\
00035056118	&PC	&2009-06-30 16:19:39	 &2009-06-30 16:44:56	  &	  1510    \\
00035056119	&PC	&2009-07-01 13:01:11	 &2009-07-01 14:50:56	  &	  1511    \\
00035056120	&PC	&2009-07-03 19:56:46	 &2009-07-03 21:50:57	  &	  2051    \\
00035056121	&PC	&2009-07-07 23:35:35	 &2009-07-07 23:46:57	  &	  672	  \\
00035056122	&PC	&2009-07-11 15:57:25	 &2009-07-11 19:21:57	  &	  2141    \\
00035056124	&PC	&2009-07-17 08:32:07	 &2009-07-17 10:16:56	  &	  1063    \\
00035056125	&PC	&2009-07-21 00:54:22	 &2009-07-21 02:06:57	  &	  1068    \\
00035056126	&PC	&2009-07-25 18:35:12	 &2009-07-25 18:54:56	  &	  1165    \\
00035056127	&PC	&2009-07-28 12:15:35	 &2009-07-28 12:30:18	  &	  864	  \\
00035056128	&PC	&2009-07-31 03:18:44	 &2009-07-31 03:31:56	  &	  783	  \\
00035056129	&PC	&2009-08-04 19:30:26	 &2009-08-04 19:46:56	  &	  943	  \\
00035056131	&PC	&2009-08-11 23:13:57	 &2009-08-11 23:29:58	  &	  944	  \\
00035056132	&PC	&2009-08-22 22:43:53	 &2009-08-22 22:56:58	  &	  761	  \\
00035056134	&PC	&2009-09-05 05:00:41	 &2009-09-05 05:17:57	  &	  1016    \\
00035056135	&PC	&2009-09-12 11:46:59	 &2009-09-12 12:07:57	  &	  1251    \\
00035056136	&PC	&2009-09-19 23:40:40	 &2009-09-19 23:59:56	  &	  1146    \\
00035056137	&PC	&2009-09-22 01:27:14	 &2009-09-22 03:16:58	  &	  1331    \\
00035056138	&PC	&2009-09-29 13:29:21	 &2009-09-29 16:51:57	  &	  1563    \\
00035056139	&PC	&2009-09-30 20:26:37	 &2009-09-30 23:43:58	  &	  947	  \\
  \noalign{\smallskip}
 \hline
 \noalign{\smallskip}
 \end{tabular}
  \end{center}
  \end{table*}

\setcounter{table}{2}

 \begin{table*}
 \begin{center}
 \caption{Observation log for IGR~J17544$-$2619. Continued. } 
 \begin{tabular}{lllll}
 \hline
 \noalign{\smallskip}
 Sequence & Instrument/Mode & Start time (UT) &  End time (UT) &  Net Exposure \\
            &     & (yyyy-mm-dd hh:mm:ss) & (yyyy-mm-dd hh:mm:ss) &  (s) \\ 
  \noalign{\smallskip}
 \hline
 \noalign{\smallskip}

00035056141	&PC	&2009-10-06 06:04:55	 &2009-10-06 06:22:58	  &	  1070    \\
00035056142     &PC   &2009-10-10 05:01:15     &2009-10-10 06:50:57	&	1571	\\
00035056143     &PC   &2009-10-13 01:56:47     &2009-10-13 02:23:56	&	1607	\\
00035056144     &PC   &2009-10-17 23:09:57     &2009-10-17 23:31:56	&	1316	\\
00035056146     &PC   &2009-10-27 03:18:01     &2009-10-27 03:33:56	&	942	\\
00035056147     &PC   &2009-10-31 05:12:03     &2009-10-31 05:30:57	&	1113	\\
00035056148     &PC   &2009-11-03 10:29:20     &2009-11-03 10:46:58	&	1058	\\
  \noalign{\smallskip}
 \hline
 \noalign{\smallskip}
 \end{tabular}
  \end{center}
  \end{table*}

\bsp

\label{lastpage}

\end{document}